\documentclass[9pt,twocolumn]{article}

\usepackage{graphicx}
\usepackage{amsmath}
\usepackage{amsfonts}
\usepackage{amssymb}
\usepackage{bm}
\usepackage{pbox}
\usepackage{bbold}

\usepackage[version=3]{mhchem}

\let\OLDthebibliography\thebibliography
\renewcommand\thebibliography[1]{
  \OLDthebibliography{#1}
  \setlength{\parskip}{0pt}
  \setlength{\itemsep}{0pt plus 0.3ex}
}

\title{Cox process representation and inference for stochastic reaction-diffusion processes}
\usepackage{authblk}
\author[1,2,3]{David Schnoerr}
\author[1,3]{Ramon Grima}
\author[2,3,*]{Guido Sanguinetti}

\affil[1]{School of Biological Sciences, University of Edinburgh, Edinburgh EH9 3JH, UK}
\affil[2]{School of Informatics, University of Edinburgh, Edinburgh EH8 9AB, UK}
\affil[3]{SynthSys, University of Edinburgh, Edinburgh EH9 3JD, UK}
\affil[*]{gsanguin@inf.ed.ac.uk}

\begin{document}

\twocolumn[
  \begin{@twocolumnfalse}
\maketitle

\begin{abstract}
Complex behaviour in many systems arises from the stochastic interactions of spatially distributed particles or agents. Stochastic reaction-diffusion processes are widely used to model such behaviour in disciplines ranging from biology to the social sciences, yet they are notoriously difficult to simulate and calibrate to observational data. Here we use ideas from statistical physics and machine learning to provide a solution to the inverse problem of learning a stochastic reaction-diffusion process from data. Our solution relies on a non-trivial connection between stochastic reaction-diffusion processes and spatio-temporal Cox processes, a well-studied class of models from computational statistics. This connection leads to an efficient and flexible algorithm for parameter inference and model selection. Our approach shows excellent accuracy on numeric and real data examples from systems biology and epidemiology. Our work provides both insights into spatio-temporal stochastic systems, and a practical solution to a long-standing problem in computational modelling.
\end{abstract}
  \end{@twocolumnfalse}
]

Many complex behaviours in several disciplines originate from a common mechanism: the dynamics of locally interacting, spatially distributed agents. Examples arise at all spatial scales and in a wide range of scientific fields,  from microscopic interactions of low-abundance molecules within cells, to ecological and epidemic phenomena  at the continental scale. Frequently, stochasticity and spatial heterogeneity play a crucial role in determining the process dynamics and the emergence of collective behaviour \cite{Bullara2015}-\cite{Cottrell2012}. 

Stochastic reaction-diffusion processes (SRDPs)  constitute a convenient mathematical framework to model such systems. SRDPs were originally introduced in statistical physics \cite{Doi1976,Doi1976b} to describe the collective behaviour of populations of point-wise agents  performing Brownian diffusion in space and stochastically interacting with other, nearby agents according to pre-defined rules. The flexibility afforded by the local interaction rules has led to a wide application of SRDPs in many different scientific disciplines where complex spatio-temporal behaviours arise, from molecular biology \cite{Takahashi2010,Bicknell2015,Grima2006}, to ecology \cite{Holmes2012}, to the social sciences \cite{Davies2013}.

Despite their popularity, SRDPs pose considerable challenges, as 
analytical computations are only possible for a handful of systems \cite{Cottrell2012}. Thus, many analytical techniques which are widely used for non-spatial stochastic systems cannot be used for SRDPs. From the practical point of view, perhaps the single most important outstanding problem is inference in SRDP models: given observations of the system, can we reconstruct the interaction rules/ local dynamic parameters? Solving this inverse problem would be important, as it would allow to quantitatively assess model fit to data and to compare different models/ hypotheses in the light of observations.

SRDPs are generally analysed by either Brownian dynamics simulations of individual particles, or by resorting to spatial discretisation, leading to the so-called  ``reaction-diffusion master equation'' (RDME) \cite{Gardiner1976, Fange2010}. The computational complexity of the RDME obviously increases as the spatial discretisation becomes finer, and in many cases the limiting process does not lead to the original SRDP \cite{Isaacson2008}. 
Significant effort has been spent to improve the performance of the two types of simulations \cite{Zon2005}-\cite{Fu2014}, however
the computational costs are still high and quickly become prohibitive for larger systems. More importantly, the lack of an analytical alternative to simulations means that evaluating the fit of a model to observations (the likelihood function) is computationally extremely expensive, effectively ruling out statistical analyses.  As far as we are aware, the few attempts at statistical inference for SRDPs either used simulation-based likelihood free methods \cite{Holmes2012}, inheriting the intrinsic computational difficulties discussed above, or abandoned the SRDP framework by adopting a coarse space discretisation  \cite{Dewar2010} or neglecting the individual nature of agents using a linear-noise approximation \cite{Ruttor2010}.

In this paper, we propose an approximate solution to the problem of computing the likelihood function in SRDPs, thus providing a principled solution to the inverse problem of model calibration. Using the classical theory of the Poisson representation (PR) for stochastic reaction processes \cite{Gardiner1977}, we show that marginal probability distributions of SRDPs can be approximated in a mean-field sense by spatio-temporal Cox point processes, a class of models widely used in spatio-temporal statistics \cite{Cressie2011}. Cox processes model the statistics of point patterns via an unobserved intensity field, a random function which captures the local mean of the observed point process. This relationship between SRDPs and Cox processes is surprising, as SRDPs are mechanistic,  microscopic descriptions of spatio-temporal systems, while Cox processes are employed phenomenologically to explain regularities in point patterns. This novel link between these two classes of models enables us to formally associate an SRDP with a continuous evolution equation on the local statistics of the process in terms of (stochastic) partial differential equations (SPDEs). Crucially, this novel representation of SRDPs allows us to efficiently approximate multiple-time marginals and thus data likelihoods associated with observed point patterns, enabling us to leverage the rich statistical literature on spatio-temporal point processes  for parameter estimation and/or Bayesian inference \cite{Cressie2011,Cseke2013a}.

\begin{figure*}[t]
\begin{center}
\centerline{\includegraphics[width=0.7\textwidth]{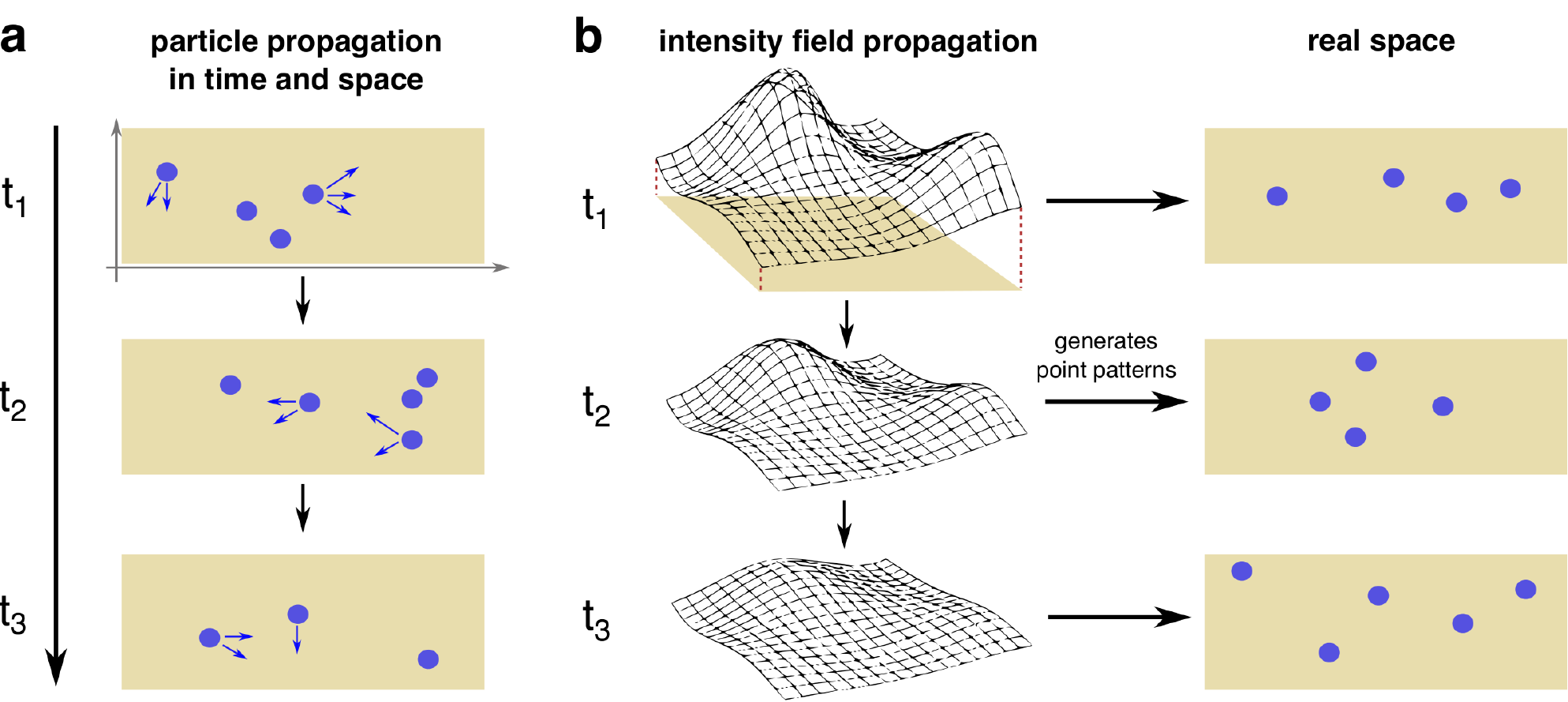}}
\caption{\textbf{Visualisation of Cox process approximation of SRDPs for multi-time points.} \textbf{(a)} Time evolution of the true SRDP in space. Particles diffuse in space, may decay or are created and react with each other. 
 \textbf{(b)} Time evolution of a Cox process. Here, the intensity field evolves in time, rather than the points in real space. The latter are merely noisy realisations of the intensity field. In particular, the points patterns at two different time points are independent of each other conditioned on the intensity field.
 }\label{fig_graphical_models}
\end{center}
\end{figure*}

We demonstrate the efficiency and accuracy of our approach for the problem of parameter inference and model selection by means of a number of numerical and real data examples using non-trivial  models from systems biology and epidemiology. Our results provide both a valuable resource for performing statistical inference and assessing model fit in this important class of models, and a novel conceptual perspective on spatio-temporal stochastic systems.

\section*{Results}
\textbf{SRDPs and the Poisson representation.}
In the classical Doi interpretation \cite{Doi1976,Doi1976b}, which we adopt here, SRDPs describe the evolution of systems of point particles performing Brownian diffusion in a spatial domain $\mathcal{D}$. While SRDPs are used in a variety of disciplines, we will use the language of chemical reactions to describe them in the following. We assume the existence of $N$ different types of particles $X_i$, or species, which can interact through a set of pre-defined rules, or reactions. We assume the structure of the model, i.e. which reactions are possible, to be known exactly; later we will relax this assumption to allow for the existence of a (finite) number of possible alternative mechanisms. Each particle of a particular species $X_i$ performs Brownian diffusion with a species-specific microscopic diffusion constant $D_i$.  Bimolecular reactions between particles occur with a certain rate whenever two particles are closer than some specified reaction range. In principle, both reaction and microscopic diffusion rate constants may be space dependent, for example to account for geometric constraints; for simplicity, we will only describe the homogeneous case here.

SRDPs are frequently analysed via coarse graining by discretising space and assuming dilute and well-mixed conditions in each compartment; in this case the dynamics in each compartment is described by the chemical master equation \cite{Gillespie2013}. Modelling diffusion of particles between neighbouring compartments as unimolecular reactions leads to the  reaction-diffusion master equation (RDME) \cite{Gardiner1976}, which describes the dynamics of a continuous-time Markov jump process.  For systems with only zeroth or first order reactions, the RDME converges to Brownian dynamics in the continuum limit. For non-linear systems, however,  this is not the case in two or more dimensions because the rate of bimolecular reactions converges to zero, independently of the scaling of the corresponding rate constant  \cite{Isaacson2008}.

Consider now a set of chemical species $X_i$ in  a finite volume divided into $L$ cubic compartments of edge length $h$ interacting via the following $R$ reactions:
\begin{align} \label{reaction_notation}
  & \sum_{i=1}^N  s_{ij} X_i^l \xrightarrow{\quad k_j \quad} \sum_{i=1}^N r_{ij} X_i^l, 
   \quad j = 1, \ldots,  R, \quad l=1, \ldots L, \\
 \label{reaction_notation_diffusion}
  & X_i^l \xrightarrow{\quad d_i \quad} X_i^m, \quad m \in \mathcal{N}(l), \quad l=1, \ldots L.
\end{align}
Here $s_{ij}$ and $r_{ij}$ are the number of reactants and product particles of species $X_i$ in the $j^{\text{th}}$ reaction, respectively, $k_j$ is the rate constant of the $j^{\text{th}}$ reaction, $X_i^m$ denotes species $X_i$ in the $m^{\text{th}}$ compartment, and $d_i$ is the diffusion rate constant of species $X_i$. The latter is related to the microscopic diffusion constant $D_i$ via $d_i=D_i/h^2$. We assume homogenous diffusion here, i.e., $d_i$ is independent of the compartment position, but it would be straightforward to extend the following analysis to space-dependent diffusion. $\mathcal{N}(l)$ denotes all the adjacent compartments of the $l^{\text{th}}$ compartment. Equation~\eqref{reaction_notation} describes chemical reactions happening in single compartments while Equation~\eqref{reaction_notation_diffusion} describes diffusion between adjacent compartments. We confine our analysis to reactions with at most two reactant and at most two product particles, i.e., $\sum_{i=1}^N s_{ir}, \sum_{i=1}^N r_{ir} \leq 2, r=1,\ldots,R$, since higher order reactions rarely occur in nature. We call a reaction linear if $\sum_{i=1}^N s_{ir},  \leq 1$ and bimolecular if $\sum_{i=1}^N s_{ir},  = 2$.
Under well-mixed and dilute conditions in each compartment, the evolution of marginal probabilities of this system is given by the RDME which is given in general form in the Methods section.  

In the case of only a single compartment, i.e.,~$L=1$, the state of the system is given by $\mathbf{n}=(n_1,\ldots, n_N)$, where $n_i$ is the number of $X_i$ particles in the system, and the time evolution of the single-time probability distribution $P(\mathbf{n},t)$ is determined by the chemical master equation (see Methods section). Gardiner derived an alternative description of the dynamics of such a system by making the ansatz of writing $P(\mathbf{n},t)$ as a Poisson mixture \cite{Gardiner1977, Thomas2010}:
\begin{align}\label{pr_ansatz}
  P(\mathbf{n},t)
  & =
    \int d \mathbf{u} \prod_{i=1}^N \mathcal{P}(n_i; u_i) p(\mathbf{u},t), \quad u_i \in \mathbb{C},
\end{align}
where $\mathbf{u}=(u_1, \ldots, u_N)$ and $\mathcal{P}(n_i;u_i)=(e^{-u_i}u_i^{n_i})/n_i!$ is a Poisson distribution in $n_i$ with mean $u_i$, and the $u_i$ are complex-valued in general. Using this ansatz in the chemical master equation one can derive an exact Fokker-Planck equation for $p(\mathbf{u},t)$ or equivalently a Langevin equation for $\mathbf{u}(t)$ \cite{Gardiner1977} (see Methods section for more details). Gardiner derived this result for the non-spatial chemical master equation and applied it to the RDME to study the corresponding continuum limit. While the PR provides an elegant analytical tool to study reaction systems, its applicability is severely hindered by the fact that the Poisson variables $u$ are in general complex-valued and hence lack a clear physical interpretation; in particular, all bimolecular reactions and all linear reactions with two non-identical product particles give rise to a complex-valued PR (for a taxonomy of which reaction systems become complex-valued see Appendix \ref{app_classification}). \\

\noindent
\textbf{Cox process representation.}
While in the classical view of the PR the auxiliary variables $u_i$ are simply introduced as a mathematical device, we can make some progress by considering a joint process over the $u_i$ and $n_i$ variables. Formally, this is equivalent to what in statistics is called  demarginalisation: a complex process is replaced by a (simpler) process in an augmented state space, such that the marginals of the augmented process return exactly the initial process. To formalise this idea, we first introduce some concepts from spatial statistics.

A (spatial) Poisson process \cite{Kingman1992} is a measure on the space of zero-dimensional subsets of a domain $\mathcal{D}$; in this work we consider Poisson processes which admit an intensity function $u(x)$, which gives the rate of finding a point in an infinitesimally small spatial region. The number of points in a finite spatial region is then a Poisson random variable with mean given by the integral of the intensity function over that region. A Cox process (also called ``doubly stochastic Poisson process'') is a generalisation of a Poisson process where the intensity field is itself a random process. Conditioned on a realisation of the intensity field, the Cox process reduces to a Poisson process (see Methods section for a more detailed definition of Poisson and Cox processes).
We will consider families of spatial Poisson (Cox) processes indexed by a time variable; importantly, in this case the intensity field can be thought of as the state variable of the system, with the actual spatial points being noisy realisations of this state (see Fig.~1 for a graphical explanation).
Our first observation follows directly from Gardiner's analysis of the continuum limit of the RDME (see Appendix \ref{app_proof} for a proof). \\

 \noindent
\textbf{Remark 1.}  Consider an SRDP on a spatial domain $\mathcal{D}$ and temporal domain $\mathcal{T}$, and let all reactions involve production or decay of at most one particle. Then, for appropriate initial conditions, $\forall t\in\mathcal{T}$ the single-time-point spatial probability distribution of the SRDP is exactly the same as of a spatial Poisson process. \\

\noindent
\textbf{General SRDPs.}
We can build on this point process representation to develop novel, mathematically consistent approximation schemes for SRDPs in general. Consider for example a bimolecular reaction of the type $A+B \xrightarrow{ k } C$ with propensity function $f(n_A,n_B) = k n_A n_B/ \Omega$, where $n_A$ and $n_B$ are the number of $A$ and $B$ particles in the system, respectively, and $\Omega$ is the system volume. While the PR for such systems is complex-valued, we can formally obtain a real system by applying a mean-field approximation that replaces the bimolecular reaction $A+B \xrightarrow{ k } C$ by the two reactions $A \xrightarrow{ k \langle n_B \rangle/ \Omega} C$ and $B \xrightarrow{ k \langle n_A \rangle / \Omega} C$ with propensity functions
$f(n_A, n_B) = k n_A \langle n_B \rangle / \Omega$ and  $f(n_A, n_B) = k n_B \langle n_A \rangle / \Omega$, respectively. Here, $\langle \cdot \rangle$ denotes the expectation of a random variable with respect to its marginal distribution.
The proposed approximation hence replaces the direct interaction of the particles by an effective interaction of $A$ with the mean-field of $B$ and vice versa. Other bimolecular reactions and linear reactions with two non-identical product particles can be approximated in a similar fashion. 
This leads to a real-valued evolution equation for the $u_i$ variables, see Methods and Appendix \ref{sec_cox_linearisation} for details and examples.

Applying this approximation to a general RDME and subsequently the PR and taking the continuum limit gives the following set of $N$ coupled SPDEs (see Methods section for a derivation)
\begin{equation} \label{pp_eq1}
\begin{split}
  d u_i(x,t) 
  & = 
    [D_i \Delta u_i(x,t) + \sum_{r=1}^R S_{ir} g_r(\mathbf{u}(x,t))] dt \\
 &   \quad
    + \sum_{r'} \sqrt {2 g_{r'}(\mathbf{u}(x,t))} dW_{r'}(x,t),
\end{split}
\end{equation}
where the sum over $r'$ runs over all reactions with two product particles of species $X_i$. In particular, this means that in the absence of reactions with two identical product molecules the diffusion term in Equation~\eqref{pp_eq1} vanishes and Equation~\eqref{pp_eq1} reduces to a partial differential equation (PDE), i.e., the $u_i(x,t)$ are deterministic. $x$ in Equation~\eqref{pp_eq1} is a spatial location, $D_i =h^2 d_i$ is the microscopic diffusion constant of species $X_i$, $\Delta$ is the Laplace operator, $\mathbf{u}(x,t) = (u_1(x,t), \ldots, u_N(x,t))$, $u_i(x,t)$ is the intensity field of species $X_i$, $dW_{r'}(x,t)$ is spatio-temporal Gaussian white noise, and we have defined the propensity functions $g_r(\mathbf{u}(x,t))$ in PR space. The latter are obtained by applying the mean-field approximation to the propensity functions $f_r(\mathbf{n})$ and subsequently replacing $n_i \to u_i (x,t)$ and $\langle n_i \rangle \to \langle u_i (x,t) \rangle$. Note that the latter denotes a local expectation of the stochastic random field $u_i (x,t)$, rather than a spatial averaging.
See Methods and Appendix \ref{sec_cox_linearisation} for more details and examples.

In order to obtain Equation~\eqref{pp_eq1} we approximated bimolecular reactions by linear reactions. Note however that the propensity functions of the latter reactions depend on the mean fields of certain species. This means that the resulting SPDEs in Equation~\eqref{pp_eq1} are generally non-linear and hence in principle able to capture non-linear dynamical behaviours.

Equation~\eqref{pp_eq1} looks similar to the spatial chemical Langevin equation \cite{Gardiner2009}, but has a different interpretation here since it describes the intensity in PR space. In particular, just as any other PDE or SPDE description in real space, the spatial chemical Langevin equation does not provide a generative model for the actual location of the events, and thus would not allow us to directly model statistically particle locations.  Notice that, as a consequence of the mean-field approximation, the mean values of the $u_i$ fields are the same as in a deterministic rate equation description; however, the dynamics of the observed variable, i.e., the points in space, remain stochastic even when the intensity field evolves deterministically.
We can therefore extend Remark 1 to obtain the following result (see Appendix \ref{app_proof} for a proof)\\

\noindent
\textbf{Result 1.} Consider the same setting as in Remark 1. Under appropriate initial conditions, if there is at least one linear reaction with two product particles of the same species, the system's single-time-point distribution is exactly the same as of a Cox process, whose intensity fulfils the  stochastic PDE (SPDE) given in Equation~\eqref{pp_eq1}.  If the system involves other types of reactions, including bimolecular reactions, the single-time probability distribution of the SRDP is approximated in a mean-field sense by that of a Poisson (Cox) process whose intensity fulfils Equation~\eqref{pp_eq1}.\\

\noindent
\textbf{The likelihood function.}
Result 1 provides an efficient means to calculate statistics such as expected number of agents within a certain volume, without the need to perform extensive Monte Carlo simulations, since it only requires to solve a (S)PDE for which a rich literature of numerical methods exists \cite{Cressie2011,Cseke2013a}. The numerical methods used in this paper are presented in the Appendix \ref{app_inference}. Importantly, we can use Result 1 to approximate the likelihood function of a configuration of points arising from an SRDP, by using the well-known Cox process likelihood:  if $u(x,t)$ is the intensity of a spatio-temporal Cox process with distribution $p(u(x,t))$ and $\mathbf{y}$ a given measurement at time $t_0$, the corresponding likelihood is given by \cite{Cressie2011}
\begin{align}\label{poisson_process_likelihood}
  p(\mathbf{y})
  & =
    \int \mathcal{D}u(x,t) \prod_{s \in \mathbf{y}} u(s, t_0) e^{- \int dx u(x, t_0)} p(u(x,t)).
\end{align}
This function can be easily optimised to yield statistical estimates of kinetic parameters from single-time observations. \\

\noindent
\textbf{Remark 2.} We would like to emphasise that in the case of a Cox process, i.e., a stochastic intensity field, the number of particles in two non-overlapping spatial regions are correlated random variables in general. The reason is that the PR ansatz in Equation~\eqref{pr_ansatz} is not merely a product of Poisson distributions, but rather an integral over such a product weighted by a corresponding mixing distribution. In the case of a Poisson process, i.e., a deterministic intensity field, in contrast, the numbers of particles in two non-overlapping spatial regions are always uncorrelated.  \\

\noindent
\textbf{Time-series observations.}
We consider next the problem of approximating the joint distribution of point patterns arising from an SRDP at different time points.
This is important when we have time-series observations, i.e., spatial measurements $\mathbf{y}=(\mathbf{y}_{t_0}, \ldots, \mathbf{y}_{t_n}), \mathbf{y}_{t_i}\subset\mathcal{D}$  at discrete time points $t_0, \ldots, t_n$, and we want to compute the likelihood $p(\mathbf{y} | \Theta)$ of the data given a model $\Theta$.   Since the system is Markovian the likelihood factorises as 
\begin{align}\label{lh_factorization}
  p(\mathbf{y} | \Theta)
  & =
    p( \mathbf{y}_{t_0} | \Theta) \prod_{i=1}^n p(\mathbf{y}_{t_i} | \mathbf{y}_{t_{i-1}}, \Theta). 
\end{align}
We would like to approximate this likelihood using the relation to Cox processes established in Result 1. While this is in principle straightforward, computing the terms $p(\mathbf{y}_{t_i} | \mathbf{y}_{t_{i-1}}, \Theta)$ involves determining the distribution over the associated $u_i(x,t)$ fields in PR space. This would involve inverting the PR transformation in Equation~\eqref{pr_ansatz}, which is computationally inconvenient. Instead, we opt for an approximation strategy: assume that we have determined the PR distribution $p(\mathbf{u}_{t_{i-1}})$ at time $t_{i-1}$, where we introduced the shorthand $\mathbf{u}_{t_{i}}=\mathbf{u}(x,t_{i})$. By definition of the intensity of a Poisson process, $\mathbf{u}
_{t_{i-1}}$ represents the expectation of the random configuration of points $\mathbf{y}_{t_{i-1}}$ at time $t_{i-1}$. We then approximate $p(\mathbf{y}_{t_i} | \mathbf{y}_{t_{i-1}}, \Theta)$ in a mean-field way by replacing the explicit dependence on $\mathbf{y}_{t_{i-1}}$ with its expectation $p(\mathbf{y}_{t_i} | \mathbf{y}_{t_{i-1}}, \Theta)\approx\langle p(\mathbf{y}_{t_i} | \mathbf{y}_{t_{i-1}}, \Theta)\rangle_{p(\mathbf{y}_{t_{i-1}}\vert \mathbf{u}_{t_{i-1}})}=p(\mathbf{y}_{t_i} | \mathbf{u}_{t_{i-1}}, \Theta)$.
Fig.~1 visualises this approximation. Fig.~1a shows the time evolution in an SRDP, while Fig.~1b shows the time evolution of a corresponding approximating Cox process. This leads to a new interpretation of the measured points $\mathbf{y}=(\mathbf{y}_{t_0}, \ldots, \mathbf{y}_{t_n})$: while they are snapshots of the actual state in the true system, they correspond to noisy realisations of the state $\mathbf{u}(x,t)$ in the Cox process picture. 
We thus have \\

\noindent
\textbf{Result 2.}
The joint $n$-time-point marginal distribution of an SRDP can be approximated in a mean-field sense by the joint probability of a Poisson (Cox) process with intensity governed by the (S)PDE in Equation~\eqref{pp_eq1}. \\

\noindent
\textbf{Relation to Gardiner's work.}
As mentioned before, Gardiner had already derived similar equations as Equation~\eqref{pp_eq1} for single-time marginals of SRDPs \cite{Gardiner1977}. Crucially, however, an approximation scheme for multiple-time joint marginals was, to our knowledge, never proposed; multi-time joint marginals are necessary for inference from time-series observations, hence the importance of Result 2. Furthermore, Gardiner's approach generally leads to a complex-valued PR; this motivates the novel approximation schemes of certain reactions that we introduced here.  It is this novel real-valued PR, together with the interpretation of the PR variables as state variables, which allows us to derive the novel relation between SRDPs and spatio-temporal Poisson (Cox) processes. \\

\begin{figure}[t]
\begin{center}
\centerline{\includegraphics[width=.3\textwidth]{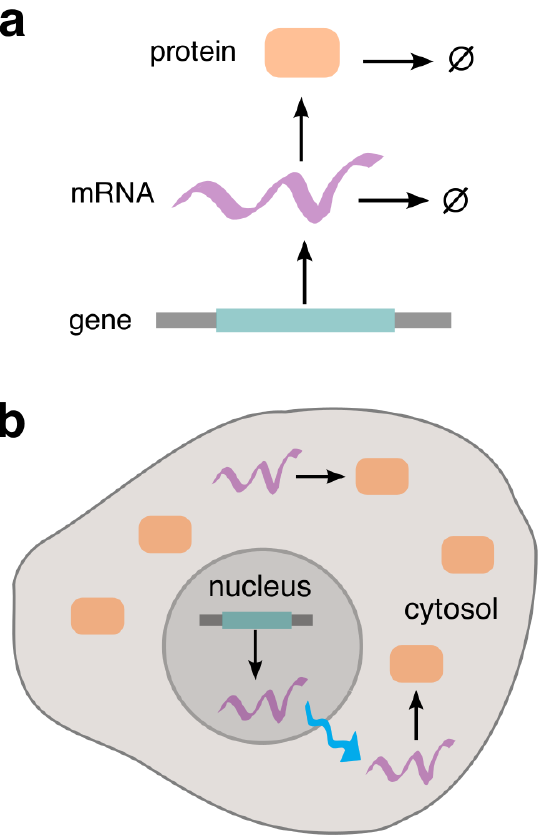}}
\caption{\textbf{Gene expression system.} \textbf{(a)} Chemical reactions taking place. \textbf{(b)} Illustration of system in a one-dimensional cell. The mRNA becomes transcribed in the nucleus, and becomes translated to proteins in the cytosol. mRNA and protein  molecules decay stochastically and undergo Brownian diffusion across the whole cell.}
\label{fig_gene_system}
\end{center}
\end{figure}

\begin{table*}[t]
\caption{\textbf{Inference results for gene expression system.} The table shows the inferred parameter values for the gene expression system illustrated in Fig.~2 with reactions in Equations \eqref{gene_system_reactions1} and \eqref{gene_system_reactions1b}. We assume measurements of the protein while the mRNA is unobserved. The inference is carried out by maximising the likelihood of simulated data for thirty measurement points separated by $\Delta t=0.5$. This procedure is carried out for hundred simulated data sets, and the mean value and standard deviation (in parenthesis) of the inferred results are displayed.}
\begin{tabular*}{\hsize}
{@{\extracolsep{\fill}}rrrrrrrrr}
&
\multicolumn1c{$r$}&
\multicolumn1c{$d_m$}&
\multicolumn1c{$d_p$}&
\multicolumn1c{$m_1$}&
\multicolumn1c{$m_2$}&
\multicolumn1c{$p_1$}&
\multicolumn1c{$p_2$}
\cr
\hline
  exact &  \multicolumn1c{$0.3$}& \multicolumn1c{$0.1$}& \multicolumn1c{$0.1$}& \multicolumn1c{$20$}& \multicolumn1c{$0.5$}& \multicolumn1c{$20$}& \multicolumn1c{$0.2$}     \cr \hline
 inferred & 0.31 (0.06) & 0.12 (0.08) & 0.14 (0.06) & 23 (12) & 0.51 (0.4) & 26 (18) & 0.25 (0.1)   \cr
\hline
\end{tabular*}\label{tab1}
\end{table*}

\begin{table*}[t]
\caption{\textbf{Inference results for gene expression system with additional autocatalytic reaction.} The table shows the inferred parameter values for the gene expression system illustrated in Fig.~2 with reactions in Equations \eqref{gene_system_reactions1} and \eqref{gene_system_reactions1b} and the additional autocatalytic reaction in Equation~\eqref{gene_system_reactions2}. Since only the difference $p_2-p_3$ is identifiable we fix $p_3=0.01$ and infer the other seven parameters. The table shows the average and standard deviations (in parenthesis) of the inference results for hundred simulated data sets.}
\begin{tabular*}{\hsize}
{@{\extracolsep{\fill}}rrrrrrrrr}
&
\multicolumn1c{$r$}&
\multicolumn1c{$d_m$}&
\multicolumn1c{$d_p$}&
\multicolumn1c{$m_1$}&
\multicolumn1c{$m_2$}&
\multicolumn1c{$p_1$}&
\multicolumn1c{$p_2$}
\cr
\hline
  exact &  \multicolumn1c{$0.3$}& \multicolumn1c{$0.1$}& \multicolumn1c{$0.1$}& \multicolumn1c{$20$}& \multicolumn1c{$0.5$}& \multicolumn1c{$20$}& \multicolumn1c{$0.2$}     \cr \hline
 inferred & 0.30 (0.05) & 0.14 (0.08) & 0.088 (0.03) & 27 (17) & 0.57 (0.3) & 24 (21) & 0.19 (0.08)   \cr
\hline
\end{tabular*}\label{tab2}
\end{table*}

\noindent
\textbf{Inference.}
Result 2 is particularly powerful statistically, because it enables us to analytically approximate the exact (intractable) likelihood $p(\mathbf{y} | \Theta)$ in Equation~\eqref{lh_factorization} by the likelihood of a spatio-temporal Cox process with intensity $u(x,t)$. The intensity itself follows the dynamics imposed by the Poisson representation in Equation~\eqref{pp_eq1}; importantly, the Poisson representation explicitly links the dynamic parameters governing the evolution of the intensity function to the microscopic diffusion and reaction rate constants of the SRDP.

Parameter estimation can therefore be performed efficiently by maximising the Cox process likelihood. In the simpler case where the intensity function evolves deterministically, the likelihood can be evaluated numerically via the solution of a system of PDEs, and the dynamic parameters can be numerically recovered using standard optimisation algorithms. In the case where the intensity function evolves stochastically, we evaluate the likelihood by an approximate filtering approach, as commonly used in many statistical and engineering applications (see Appendix \ref{app_inference} for algorithmic details). 

The availability of a likelihood function enables us to provide a statistically meaningful, data-driven assessment of how well a model describes the data. This is particularly important when there is uncertainty as to the precise mechanism underlying the data, e.g. the exact reactions or species involved. Likelihood estimates, appropriately penalised to account for model complexity, can then be used to select models according to their support from the data. 

It is important to notice that our approach directly optimises the kinetic parameters of the model, rather than fitting an intensity function to the observed points and then fitting the dynamics. Since kinetic parameters are usually much fewer than the number of observations available, the risk of over-fitting is generally low in our approach.

Next, we apply Result 1 and Result 2 to several examples, and perform parameter inference by maximising the data likelihood. We solve the corresponding (S)PDEs numerically by projecting them onto a finite set of spatial basis functions, see Appendix \ref{sec_basis_proj} for details. \\

\begin{figure*}[t]
\begin{center}
\centerline{\includegraphics[width=0.9\textwidth]{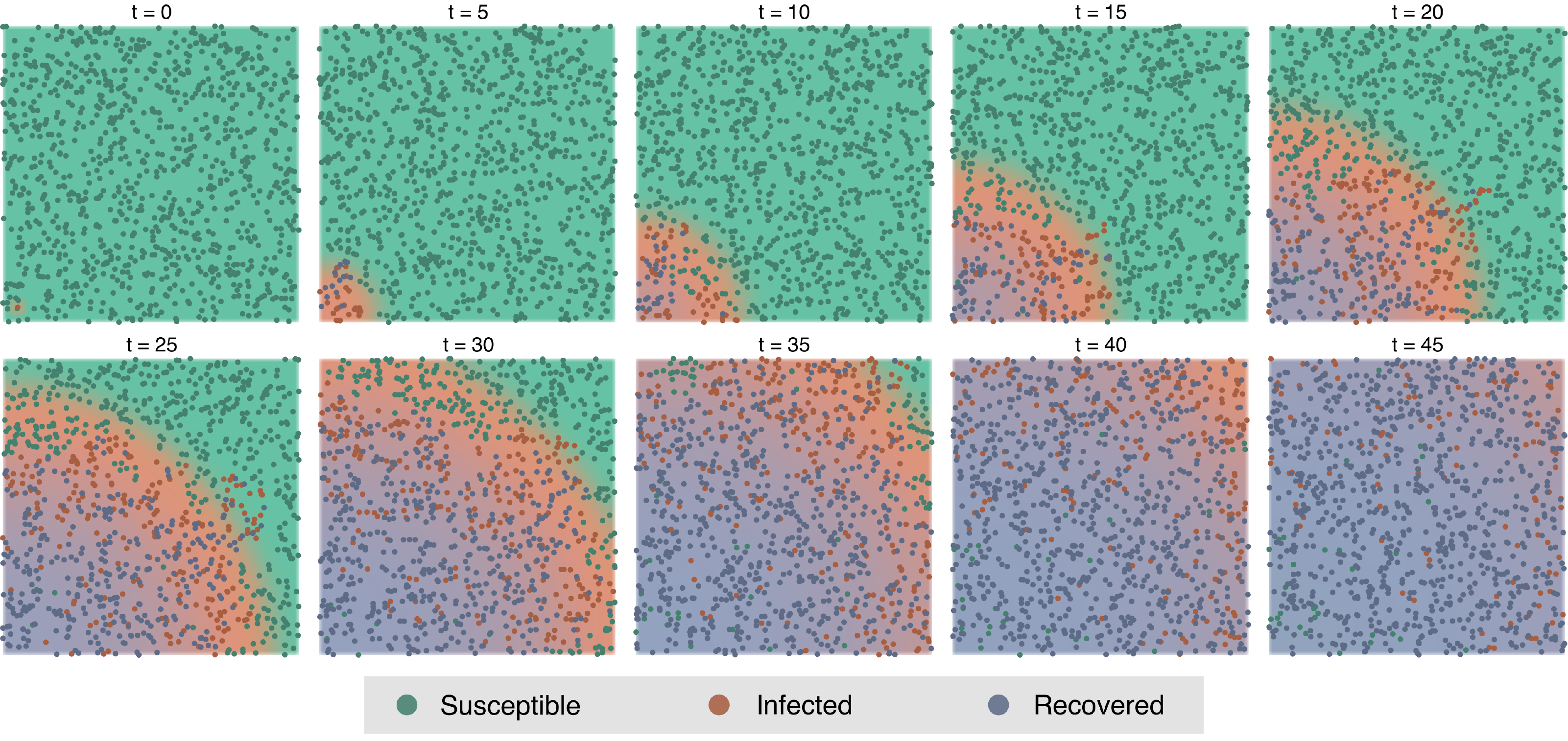}}
\caption{\textbf{Dynamics of SIRS system.} The figure shows the time evolution of a single simulated realisation (points) and of the prediction of our method (background colours) for the SIRS system with reactions in Equation~\eqref{sirs} for time $t=0$ to $t=45$ with steps of $\Delta t = 5$. For the simulation we use the parameters $(S_{\text{ini,}} k, w, d, r, s) = (10^3, 10^4, 0.02, 0.0002, 0.3, 0.01)$  and for the point process prediction the corresponding inferred parameters. The background is an RGB image with the three colour components being proportional to the intensity fields of the three species $S$ (turquoise), $I$ (bronze) and $R$ (blue).  Notice how the mean-field approximation captures the complex behaviour of a wave of infection spreading through the domain from the bottom left corner.}\label{fig_sir_time}
\end{center}
\end{figure*}

\noindent
\textbf{Parameter estimation for a gene expression system.}
To demonstrate the accuracy of our method, we first consider simulated time-series data in this section.
Consider a gene expression system as illustrated in Fig.~2. For simplicitly, we consider a one-dimensional version of this system here with the nucleus located at one side of the cell.
A gene located in the nucleus is transcribed into mRNA molecules. The latter decay and diffuse across the whole cell and are translated into proteins in the cytosol. The protein molecules also diffuse across the whole cell and decay. For simplicity, we do not model the gene explicitly but assume that mRNA becomes transcribed with a certain fixed rate $m_1$ homogeneously in the nucleus. The corresponding reactions are 
\begin{align}\label{gene_system_reactions1}
  & \varnothing \xrightarrow{\quad m_1 \quad} M, \quad
  M \xrightarrow{\quad m_2 \quad} \varnothing, \\
  & \label{gene_system_reactions1b}
  M \xrightarrow{\quad p_1 \quad} M + P, \quad P \xrightarrow{\quad p_2 \quad} \varnothing,
\end{align}
where $M$ and $P$ denote the mRNA and protein, respectively. 
For this system, the SPDE of our method in Equation~\eqref{pp_eq1} becomes deterministic and thus corresponds to a Poisson process.

In addition to the reaction parameters $m_1, m_2, p_1$ and $p_2$, we have to infer the nucleus size $r$, as well as the diffusion rates $d_m$ and $d_p$ of the mRNA and protein, respectively, summing up to a total number of seven parameters. 
We assume that the positions of the protein molecules are observed at thirty time points, while the mRNA is unobserved. 
The results for one parameter set are shown in Table 1. Considering that we observe the protein at only thirty time points with unobserved mRNA and that we have seven unknown parameters, the inferred average values are remarkably close to the exact values. Moreover, the standard deviations of the inferred results for single data sets are small, demonstrating the accuracy and precision of our method.

Next, we extend the system in Fig.~2 by adding an autocatalytic  reaction for the protein,
\begin{align}\label{gene_system_reactions2}
  P \xrightarrow{\quad p_3 \quad} P+P.
\end{align}
Including this reaction leads to a non-vanishing noise term in Equation~\eqref{pp_eq1} and the system corresponds to a Cox process. We note that the system has a steady state only if $p_3<p_2$ , with an otherwise exponentially growing mean protein number. On the mean level only the difference $p_2-p_3$ is identifiable, and we fix $p_3=0.01$. We thus infer the same parameters as in the previous case, but this time modelled as a Cox process. Table 2 shows the results indicating the accuracy of our method. See Appendix \ref{sec_example_gene} for more information on the used equations and algorithmic details. \\

\noindent
\textbf{Parameter estimation for an SIRS model.} 
We next consider the SIRS system, a popular model for describing the dynamics of an infection spreading through a population. Such systems are frequently modelled as SRDPs \cite{Peruani2013} or discretised versions thereof \cite{Abdullah2015}. We consider a system in the two-dimensional square $[0,1] \times [0,1]$. The system comprises a susceptible (S), an infected (I) and a recovered species (R), which perform Brownian diffusion and interact via the reactions 
\begin{align}\label{sirs}
  S+I \xrightarrow{\quad k, w \quad} 2I, \quad
  I \xrightarrow{\quad r \quad} R,  \quad
  R \xrightarrow{\quad s \quad} S,
\end{align}
where the bimolecular infection is characterised by the microscopic reaction rate $k$ and the reaction range $w$.  We assume that all three species diffuse with the same diffusion rate $d$. We assume further that initially there are no recovered (R) particles,  $S_{\text{ini}}$ susceptible (S) particles placed uniformly over the whole area, and one infected (I) particle located at $[0.05,0.05]$.  We consider the case that all three species are observed and perform inference for forty simulated data points using Result 2,  thereby replacing $k$ and $w$ by an effective bimolecular reaction parameter $k^{\text{PR}}$. The model thus has four parameters that need to be inferred: the diffusion rate $d$, the recovery rate $r$, the susceptible rate $s$ and the bimolecular infection rate $k^{\text{PR}}$. Table 3 shows the corresponding results, demonstrating the accuracy and precision of our method. The computational efficiency of our method in comparison to stochastic simulations is particularly pronounced here. For the first parameter set in Table 3, for example, the Brownian dynamics simulation of a single realisation of the system takes about 250 seconds, while the whole inference procedure for the four parameters Appendix \ref{sec_example_sirs} for more details.

\begin{table*}[t]
\caption{\textbf{Inference results for SIRS system.} The table shows the results for parameter inference for the SIRS system with reactions given in Equation~\eqref{sirs}. The inference is carried out by maximising the likelihood of simulated data for forty measurement points. This procedure is carried out for two hundred simulated data sets, and the mean value and standard deviation (in parenthesis) of the inferred results are displayed.}
\centering
\begin{tabular}
{@{\extracolsep{\fill}}rrrrrrrr}
\multicolumn1c{}&
\multicolumn1c{$10^{3} \times d$}&
\multicolumn1c{$10 \times r$}&
\multicolumn1c{$10 \times s$} &
\multicolumn1c{$10^3 \times k^{\text{PR}}$} &
\multicolumn1c{$k$ } &
\multicolumn1c{$w$ } &
\multicolumn1c{$S_{\text{ini}}$ }
\cr
\hline
   exact & \multicolumn1c{1} & \multicolumn1c{0.2} & \multicolumn1c{2} & \multicolumn1c{-}  & 100 & 0.01 & 200 \cr
   inferred & 0.8 (0.3) & 0.19 (0.09) & 1.8 (1.2) & 2.5 (0.5) & - & -  \cr
\hline
   exact & \multicolumn1c{1} & \multicolumn1c{0.2} & \multicolumn1c{2} & \multicolumn1c{-}  & 100 & 0.01 & 300 \cr
   inferred & 0.9 (0.4) & 0.15 (0.06) & 1.4 (0.9) & 2.4 (0.5) & - & -  \cr   
\hline
   exact & \multicolumn1c{1} & \multicolumn1c{2} & \multicolumn1c{2} & \multicolumn1c{-} & 100 & 0.02 & 200 \cr
   inferred & 1.0 (0.6) & 1.6 (0.7) & 1.5 (1.0) & 3.4 (1.1) & - & -  \cr  
\hline
   exact & \multicolumn1c{1} & \multicolumn1c{0.2} & \multicolumn1c{2} & \multicolumn1c{-} & 1000 & 0.005 & 200 \cr
   inferred & 0.8 (0.4) & 0.21 (0.11) & 2.2 (1.6) & 2.4 (0.5) & - & -   \cr
\hline
   exact & \multicolumn1c{2} & \multicolumn1c{0.2} & \multicolumn1c{2} & \multicolumn1c{-} & 100 & 0.01 & 100 \cr
   inferred & 1.6 (0.8) & 0.19 (0.09) & 1.9 (1.2) & 4.6 (1.1) & - & -   \cr
\hline
\end{tabular}\label{tab3}
\\
\vspace{0.5cm}
\noindent\rule{18cm}{0.4pt}
\vspace{0.5cm}
\end{table*}

Fig.~3 visualises the dynamics of the SIRS system for one parameter set. Individual points from a simulation are shown in different colours (turquoise for S, bronze for I and blue for R), while the background RGB colours represent a superposition of the respective intensity fields with optimised parameters. Notice how the PR approximation is able to capture the emerging behaviour of a wave of infection sweeping through the domain from bottom left to top right, before the establishment of a dynamic equilibrium between the three population. Such a phenomenon is clearly due to the spatial aspect of the system, and could not have been recovered using an inference method that does not incorporate spatial information. Indeed, the overall number of infected individuals rises rapidly and remains essentially constant between time 20 and time 35 before dropping to steady state, a behaviour which is simply not possible in a non-spatial SIRS model. \\

\begin{figure*}[t]
\begin{center}
\centerline{\includegraphics[width=0.8\textwidth]{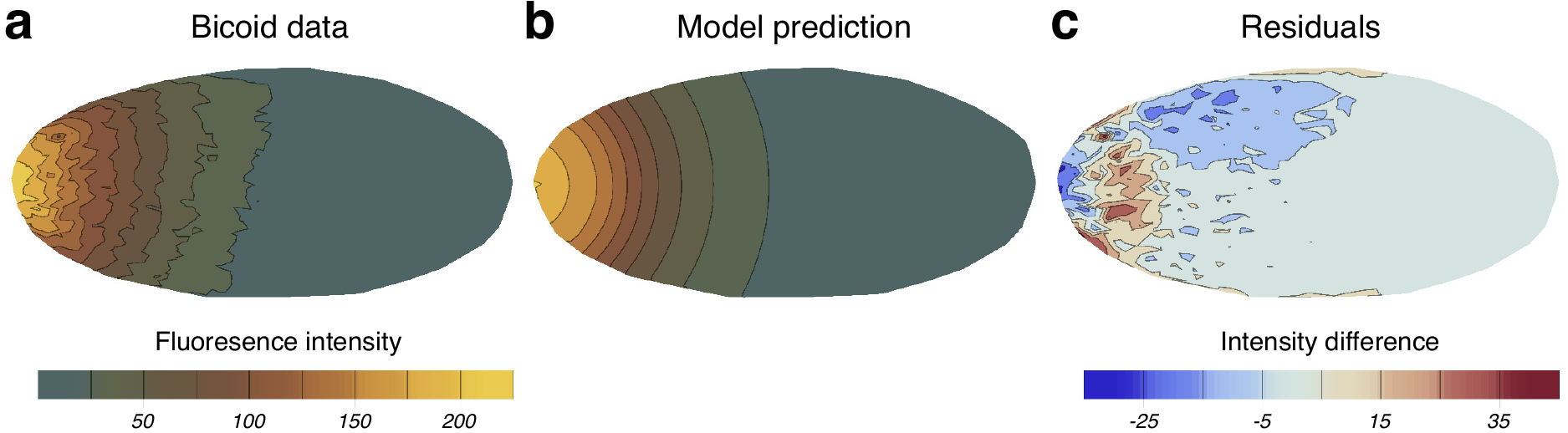}}
\caption{\textbf{Results for the Drosophila embryo Bicoid data.}
\textbf{(a)} Measurement data of Bicoid fluorescent intensity across a single embryo. \textbf{(b)} Corresponding prediction of our point process model. \textbf{(c)} Difference of the experimental data and point process prediction. We observe the point process prediction agrees well with the experimental data. The point process prediction is obtained by solving Equation~\eqref{pp_eq1} numerically for the inferred parameter values maximising the data likelihood.}
\label{fig_bicoid1}
\end{center}
\end{figure*}

\noindent
\textbf{Parameter estimation for Drosophila embryo data.}
Finally, we apply our method to real gene expression data for the Bicoid protein at cleavage stage 13 in the Drosophila embryo. The data for seventeen embryos can be obtained from the FlyEx  database \cite{Flyex1}.  The data consists of fluorescence intensity measurements on a spatial grid and is shown for one embryo in Fig.~4a. The protein becomes expressed in some region at the left end of the embryo and then diffuses across the embryo and decays. The system is typically modelled by a linear birth-death process \cite{Dewar2010, Ruttor2010}, and we assume the protein to be expressed within a certain distance $r$ from the left end of the embryo. At cleavage stage 13 the system is supposed to be in steady state and we can perform inference using Result 1 and Equation~\eqref{pp_eq1}. For simplicity, we project the data to one dimension (see Appendix \ref{sec_example_drosophila} for more details).

The system has four parameters: the creation range $r$, the diffusion rate $d$, the production rate $c_1$ and decay rate $c_2$ of the Bicoid protein. For steady-state data not all parameters but only certain ratios are identifiable. We thus infer the creation range $r$, the diffusion rate $d$ and the ratio $c=c_1/c_2$. For the average of the inferred parameters and their standard deviations (shown here in parentheses) across the ensemble of embryos we obtain
\begin{align}\label{bicoid_inferred_pars}
  r = 0.26 (0.09), \quad d =  0.023 (0.005), \quad c = 1.3 (0.2) \times 10^4,
\end{align}
with standard deviations of about $20\%$ to $30\%$. We find that these results do not change significantly under variations of the initial parameter values used in the likelihood optimiser.

Fig.~4 illustrates the inference result for one embryo. Figs. 4a and 4b show the Bicoid density across the whole embryo for experimental data and the PR prediction, respectively. We observe good agreement between the measurement data and the point process approximation. Fig.~4c shows a plot of the model residuals (difference between model predictions and real data); as can be seen, these are generally comparatively small. As could be expected, the larger errors are concentrated around the steeply changing gradient region between the anterior segments and the main body of the embryo. \\

\begin{figure*}[t]
\begin{center}
\centerline{\includegraphics[width=0.7\textwidth]{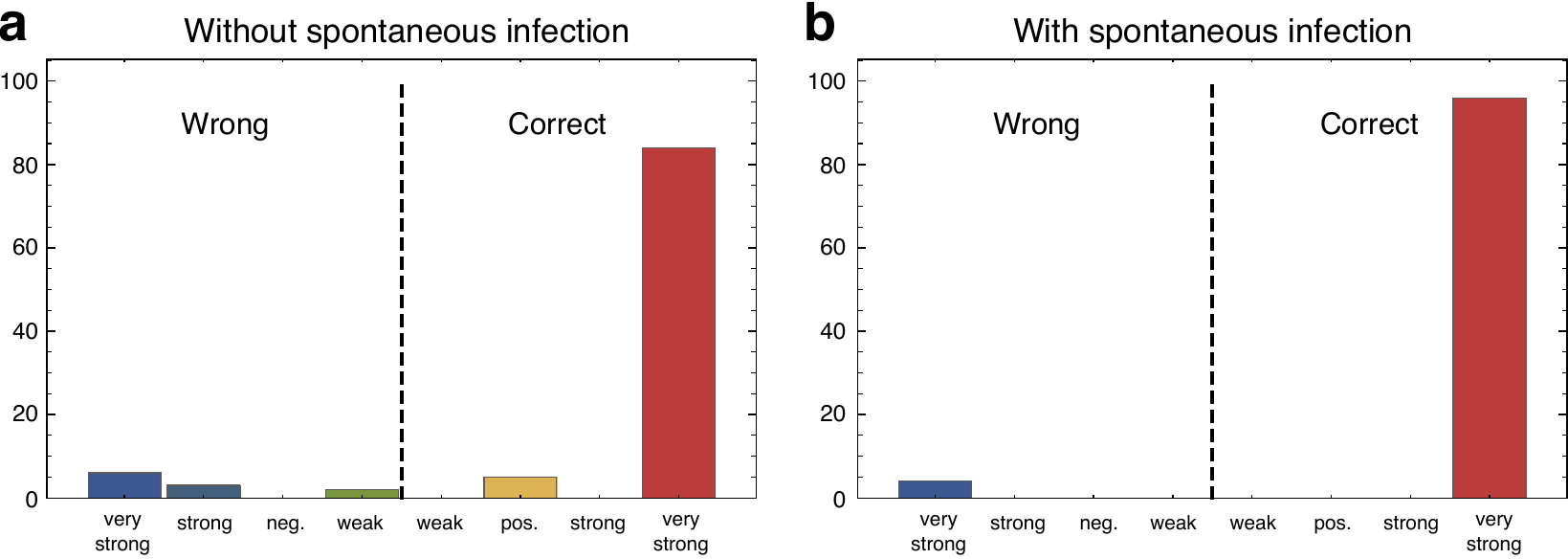}}
\caption{\textbf{Model selection results for SIRS system.} We use the Bayesian information criterion (BIC) for model selection for the SIRS system with reactions in Equation~\eqref{sirs} and the additional spontaneous infection reaction in Equation~\eqref{sirs2}. The sign of the difference in the BIC numbers of the two models determines if the correct model is selected, and the corresponding magnitude how confident this choice is. We simulated twenty experiments for five parameter sets each. The figures show the combined results of these hundred experiments. 
\textbf{(a)} The true system used to generate the data does not include spontaneous infection. The parameter sets are the same as in Table \ref{tab3}. \textbf{(b)} The true system used to generate the data does include spontaneous infection. The parameter sets are the same as in Table \ref{tab3} but with modified bimolecular infection rate which we set to $k=10, 5, 10, 100$ and $10$, respectively. The spontaneous infection rate is set to $v=0.002$ for all parameter sets. In both cases, we observe that our method selects the correct model in more than $80\%$ of the cases, and in most of these cases with ``very strong'' confidence. This demonstrates the strong performance of our method for this model-selection problem.
 }\label{fig_model_comparison}
\end{center}
\end{figure*}

\noindent
\textbf{Model selection for an SIRS model.}
Next, we use Result 2 to perform model selection. Specifically, we use our method to decide which of two given microscopic models is more likely to be the true model underlying some given data set. To this end, we use the Bayesian information criterion (BIC) \cite{Schwarz}. The BIC for a model is the negative log-likelihood penalised by a term depending on the number of inferred parameters and number of measurement points. The model with the lower BIC is then chosen to be the true model. 

As an example, we modified the SIRS model of Equation~\eqref{sirs} by including the possibility of a spontaneous, spatially homogeneous infection of susceptible agents, according to the reaction
\begin{align}\label{sirs2}
  S \xrightarrow{\quad v \quad} I.
\end{align}
We consider two scenarios: the true microscopic model used to generate the data does or does not contain the spontaneous infection reaction. In either case, we use our method to select the true model. To this end, we optimise the likelihood with respect to the parameters using our method for both models, and compare the corresponding BICs. Figs.~5a and 5b show the results for the two scenarios of true model without and with spontaneous infection, respectively. The figures show how often our method selected the right or wrong model, and with which confidence level. Each of the figures shows the combined results for five different parameter sets and twenty independent simulations for each parameter set. We find that our method chose the correct model in the vast majority of the cases ($89 \%$  where the true system does not include spontaneous infection and $96 \%$ where the true system does contain spontaneous infection.). Moreover, our method choses the correct model with a ``very strong'' confidence in most of the cases. This shows that our method is well suited for the problem of model selection. 
The effectiveness of our model selection approach is remarkable, since the two mechanisms (spontaneous infection and contact infection) can lead to identifiability problems. Such problems are particularly acute when spatial heterogeneities even out rapidly, as in the case of fast diffusion: the few mistakes that our model selection approach makes are primarily due to random samples of the SRDP when the infection spreads particularly fast, so that, for most time points, the process is effectively equilibrated.

\section*{Discussion}

We considered two popular classes of models for studying stochasticity in spatio-temporal systems; stochastic reaction-diffusion processes (SRDPs) and spatio-temporal point processes. The two classes of models are both commonly used in many disciplines such as epidemiology \cite{Davies2013,Grell} and social sciences \cite{Zammit}, however they are widely perceived as conceptually distinct. SRDPs are microscopic, mechanistic descriptions used to predict the dynamics of spatially interacting particles, whereas point processes are typically used empirically to perform inference tasks for systems for which no microscopic description exists. The two approaches therefore seem to be orthogonal to each other. 

However, in this paper we have shown that the two methods are intimately related. By using the Poisson representation (PR) we established a Cox process representation of SRDPs, which is exact for certain classes of systems and approximate for others. This novel representation enables us to apply a wide range of statistical inference methods to SRDPs, which has not been possible so far. We applied the developed method to several example systems from systems biology and epidemiology and obtained remarkably accurate results.

Since our method agrees with a deterministic rate equation description on the mean level, bimolecular reactions may lead to deviations from the true mean, which is known to be the case in some non-spatial scenarios \cite{Ramaswamy2012}. Since in our method distributions are given as real Poisson mixtures, sub-Poissonian fluctuations cannot be captured. 
However, Gardiner showed that fluctuations of SRDPs are dominated by diffusion on small length scales and therefore Poissonian \cite{Gardiner2009}, which may explain the accuracy of our method.

Most inference methods in the literature for SRDPs are either based on Brownian dynamics simulations or  stochastic simulations of spatially discretised systems using the RDME. Both approaches are computationally extremely expensive and quickly become unfeasible for larger systems and in particular for inference purposes. In contrast, our method relies on the solution of (S)PDEs for which a rich literature of efficient numerical methods exists. For the studied example systems our method turned out to be highly efficient: the computational time for inferring four unknown parameters for the SIRS system, for example, was found to be of the order of $10$ seconds on a $3.1$GHz processor. We therefore expect our method to be applicable to significantly larger systems containing more species and unknown parameters. 
Remarkably, simulating a single realisation of the SIRS system from Brownian dynamics simulations took about an order of magnitude longer than the whole inference procedure using our method, i.e., optimising the likelihood with respect to the parameters, indicating the immense computational costs of inference methods based on such simulations. 

Having access to a likelihood function also provides a major advantage in handling model uncertainty: our results on a spatial SIRS model show that likelihood-based criteria can efficiently and accurately discriminate between competing models. This success raises the possibility that our approach could lay the foundations for structure learning of spatio-temporal stochastic systems: leveraging spatially resolved data not only to identify parameters, but to learn directly the mechanisms underlying the data. The availability of a likelihood approximation makes it in principle possible to borrow techniques from fields where structure learning is more established and where efficient network learning algorithms based on regularised regression or random forests are routinely used, such as learning gene regulatory networks \cite{Bonneau06,Huynh-Thu15}.

Our approach can also readily handle spatial heterogeneity in the reaction or diffusion rates: the gene regulation example showed that the method can precisely identify simple geometric features of the system, such as the radius of the cell nucleus. While our examples are primarily illustrative of the methodology, and hence simple, it would be in principle straightforward to generalise the approach to SRDPs defined on complex geometries, such as the intracellular landscapes revealed by X-ray tomography \cite{Do15}. Learning complex geometries directly from data would potentially be more challenging, however, as it would generally require learning a large number of parameters.

While we believe that the derived representation of SRDPs in terms of Cox processes brings clear advantages from a statistical point of view, it is also important to acknowledge the limitations implied by the employed mean-field approximations. Perhaps the most important step in our approximation is the mean-field treatment of multi-time joint distributions in Result 2. As noticed before, this replaces direct dependencies between particle locations at different time points with indirect dependences through the intensity field. This implies that self-excitatory behaviours, such as clustering, cannot be accurately captured; at best, these will be mimicked by a local increase in the intensity field within a Cox process framework. More complex point processes that can account for self-excitatory behaviour do exist \cite{Hawkes74}; in our opinion, it is a question of considerable theoretical interest whether such processes can also arise from a dynamical SRDP representation.

{\small

\section*{Methods}

\textbf{The chemical master equation.}
Consider a system of $N$ species $X_i, i = 1,...,N$ that interact stochastically via $R$ reaction channels
\begin{align} \label{reaction_notation2}
  \sum_{i=1}^N  s_{ij} X_i \xrightarrow{\quad k_j \quad} \sum_{i=1}^N r_{ij} X_i, 
   \quad j = 1, \ldots,  R,
\end{align}
where $k_j$ is the rate constant of the $j^{\text{th}}$ reaction and the $s_{ij}$ and $r_{ij}$ are the non-negative integer numbers.  Define the stoichiometric matrix $\mathbf{S}$ as $S_{ij}  =  r_{ij} - s_{ij}$; the $j^{\text{th}}$ reaction is of order $m$ if $\sum_{i=1}^N  s_{ij}=m$. We only consider reactions satisfying $\sum_{i=1}^N s_{ij}, \sum_{i=1}^N r_{ij} \leq 2$,
i.e., reactions with a maximum of two reactant and a maximum of two product particles, since higher order reactions rarely occur in nature.
Denote as $\mathbf{n}=(n_1, \ldots, n_N)$ the state of the system, where $n_i$ is the copy number of species $X_i$. Under well-mixed and dilute conditions, the time evolution of the (single-time) marginal probability distribution of the system obeys the chemical master equation \cite{Gillespie2013}
\begin{align}\label{cme}
  \partial_t P(\mathbf{n},t) 
  & = 
    \sum_{r=1}^R f_r (\mathbf{n} - \mathbf{S}_r) P(\mathbf{n} - \mathbf{S}_r, t) 
    - \sum_{r=1}^R f_r (\mathbf{n}) P(\mathbf{n}, t),
\end{align}
where $\mathbf{S}_r$ is the $r^{\text{th}}$ column of the stoichiometric matrix $\mathbf{S}$. The propensity function $f_r (\mathbf{n}) dt$ gives the probability for the $r^{\text{th}}$ reaction to happen in an infinitesimal time interval $dt$ and is given by
\begin{equation}\label{general_propensity}
f_r(\mathbf{n}) = k_r \Omega \prod_{k=1}^N \frac{n_k!}{(n_k - s_{kr})! \Omega^{s_{kr}}}.
\end{equation}
Here, $\Omega$ is the volume of the system.\\

\noindent
\textbf{The Poisson representation.}
The Poisson representation makes the ansatz to write $P(\mathbf{n},t)$ as a Poisson mixture \cite{Gardiner1977}
\begin{align}\label{pr_ansatz2}
  P(\mathbf{n},t)
  & =
    \int d \mathbf{u} \ \mathcal{P}(n_1; u_1) \ldots \mathcal{P}(n_N; u_N)  p(\mathbf{u},t), \quad u_i \in \mathbb{C},
\end{align}
where $\mathbf{u}=(u_1, \ldots, u_N)$ and $\mathcal{P}(n_i;u_i)=(e^{-u_i}u_i^{n_i})/n_i!$ is a Poisson distribution in $n_i$ with mean $u_i$, and the $u_i$ are complex in general.
The integrals in Equation \eqref{pr_ansatz2} in general run over the whole complex plane for each $u_i$. 
Using the ansatz \eqref{pr_ansatz2} in the generating function equation which can be derived from Equation \eqref{cme} one can derive the following PDE for the distribution $p(\mathbf{u},t)$ \cite{Gardiner1977},
\begin{equation}\label{pr_fpe_general}
\begin{split}
  \partial_t p(\mathbf{u},t)
  & = 
    \sum_{r=1}^R \Omega k_r  \left( \prod_{i=1}^N \left(1-\frac{\partial}{\partial u_i}\right)^{r_{ir}}
    - \prod_{i=1}^N \left( 1 - \frac{\partial}{\partial u_i}\right)^{s_{ir}} \right) \\
    & \quad \times \prod_{j=1}^N \Omega^{-s_{jr}}  u_j^{s_{jr}} p(\mathbf{u},t).
\end{split}
\end{equation}
Note that this PDE generally involves derivatives of higher order than two, which means that $p(\mathbf{u},t)$ can generally become negative in which case it does not admit a probabilistic interpretation. However, since we only consider reactions satisfying $\sum_{i=1}^N s_{ir}, \sum_{i=1}^N r_{ir} \leq 2$, 
Equation \eqref{pr_fpe_general} simplifies to
\begin{equation}\label{pr_fpe_reduced}
\begin{split}
  \partial_t p(\mathbf{u},t)
  & = 
    - \sum_{i=1}^N \frac{\partial}{\partial u_i} \left[ A_i (\mathbf{u}) p(\mathbf{u}, t) \right] \\
  & \quad
    + \frac{1}{2} \sum_{i,j=1}^N \frac{\partial}{\partial u_i} \frac{\partial}{\partial u_j}
     \left[ B_{ij}(\mathbf{u}) p(\mathbf{u}, t) \right],
\end{split}
\end{equation}
which is a Fokker-Planck equation (FPE) with drift vector $\mathbf{A}(\mathbf{u})$ and diffusion matrix $\mathbf{B}(\mathbf{u})$ given by
\begin{align}
  A_i(\mathbf{u})
  & = 
    \sum_{r=1}^R S_{ir} g_r(\mathbf{u}), \\
\label{pr_diff}
  B_{ij}(\mathbf{u})
  & =
    \sum_{r=1}^R g_r(\mathbf{u}) (r_{ir} r_{jr} - s_{ir} s_{jr} - \delta_{i,j} S_{ir}), \\
\label{pr_prop}
  g_r(\mathbf{u})
  & = 
    \Omega k_r \prod_{j=1}^N \Omega^{-s_{jr}} u_j^{s_{jr} },
\end{align}
where $\delta_{i,j} $ denotes the Kronecker delta.
The corresponding Langevin equation reads
\begin{align}\label{pr_cle_reduced}
  d \mathbf{u}
  & = 
    \mathbf{A}(\mathbf{u}) dt + \mathbf{C}(\mathbf{u}) d \mathbf{W}, \quad \mathbf{C} \mathbf{C}^T = \mathbf{B},
\end{align}
where $d \mathbf{W}$ is a $l$-dimensional Wiener process and $l$ is the number of columns of $\mathbf{C}$.

Depending on the reactions in the system, the diffusion matrix may be zero, in which case the Langevin equations in \eqref{pr_cle_reduced} reduce to deterministic ordinary differential equations. On the other hand, depending on the reactions, $\mathbf{B}(\mathbf{u})$ is not positive-semidefinite and thus $\mathbf{C} \mathbf{C}^T = \mathbf{B}$ cannot be fulfilled for real $\mathbf{u}$, 
which means that Equation \eqref{pr_fpe_reduced} is not a proper FPE in real variables. Rather, it needs to be extended to complex space. Specifically, this is the case whenever the system contains bimolecular reactions or reactions with two non-identical product molecules.

An important property of the PR is that the mean values of the particle numbers $n_i$ are equal to the mean values of the corresponding PR variables $u_i$, i.e., $\langle n_i \rangle=\langle u_i \rangle$. \\

\noindent
\textbf{The reaction-diffusion master equation.}
Consider a system as in Equation \eqref{reaction_notation2} but in an $M-$dimensional volume discretised into $L$ cubic compartments of edge length $h$ and volume $h^M$.
Denote as $\mathbf{n}=(n_1^1, \ldots, n_N^1, \ldots, n_1^L, \ldots, n_N^L)$ the state of the system, where $n_i^l$ is the copy number of species $X_i$ in the $l^{\text{th}}$ compartment. Under well-mixed and dilute conditions in each compartment, the reaction dynamics in each compartment is governed by a corresponding chemical master equation as in Equation \eqref{cme}. If we model diffusion of species $X_i$ between neighbouring compartments by linear reactions with rate constant $d_i=D_i /h^2$, where $D_i$ is the microscopic diffusion constant of species $X_i$, the time evolution of the (single-time) marginal probability distribution of the system obeys the RDME \cite{Gillespie2013}:
\begin{align}\label{rdme1}
  \partial_t P(\mathbf{n},t) 
  & = 
  \sum_{l=1}^L \sum_{m \in \mathcal{N}(l)} \sum_{i=1}^N d_i [ (n_i^m+1)P(\mathbf{n} + \boldsymbol{\delta}_i^m - \boldsymbol{\delta}_i^l, t) 
  - n_i^l P(\mathbf{n}, t)]
  \\
  \label{rdme2}
  & \quad
   +\sum_{l=1}^L \sum_{r=1}^R [ f_r (\mathbf{n}^l - \mathbf{S}_r) P(\mathbf{n} - \mathbf{S}_r^l, t) 
    -  f_r (\mathbf{n}^l) P(\mathbf{n}, t)],
\end{align}
where $f_r (\mathbf{n}^l)$ is the propensity function of the $r^{\text{th}}$ reaction evaluated at the state vector $\mathbf{n}^l=(n_1^l, \ldots, n_N^l)$ of the $l^{\text{th}}$ compartment, $\boldsymbol{\delta}_i^l$ is a vector of length  $N*L$ with the entry corresponding to species $X_i$ in the $l^{\text{th}}$ compartment equal to $1$ and all other entries zero and $\mathbf{S}_r^l$ is a vector of length $N*L$ with the entries corresponding to the $l^{\text{th}}$ compartment equal to the $r^{\text{th}}$ row of the stoichiometric matrix $\mathbf{S}$ and zero otherwise. \\

\noindent
\textbf{Real-valued Poisson representation in space.}
We next apply the PR to the RDME in Equations \eqref{rdme1} and \eqref{rdme2} after applying the mean-field approximations defined in the Results section to bimolecular reactions and reactions with two non-identical product molecules, and subsequently take the continuum limit. Consider first the diffusion term in Equation \eqref{rdme1}. Since different species do not interact with each other if there are no chemical reactions happening, we can consider a system containing only a single species, say species $X_1$, for which Equation \eqref{rdme1}  reduces to
\begin{align}\label{only_diffusion1}
  \partial_t P(\mathbf{n},t) 
  & = 
  \sum_{l=1}^L \sum_{m \in \mathcal{N}(l)} d [ (n^m+1)P(\mathbf{n} + \boldsymbol{\delta}^m - \boldsymbol{\delta}^l, t) 
  - n^l P(\mathbf{n}, t)],
\end{align}
where $\mathbf{n}=(n^1,\ldots, n^L)$, $n^m$ is the number of $X_1$ particles in the $m^{\text{th}}$ compartment, $\boldsymbol{\delta}^m$ is a vector with a one in the $m^{\text{th}}$ entry and zero otherwise, $d$ is the diffusion constant of species $X_1$, and the sum over $m$ runs over all neighbouring compartments $\mathcal{N}(l)$ of the $l^{\text{th}}$ compartment. For this system the PR is real and deterministic, and we use the PR without any approximations. 
The corresponding Langevin equation reads
\begin{align}\label{diffusion_rdme_pr}
  d u^l 
  & =
    D \frac{2 M u^l - \sum_{m \in \mathcal{N}(l)} u^m}{h^2} dt, \quad l=1,\ldots, L,
\end{align}
where $M$ is the spatial dimension of the system and $D= d h^2$ the microscopic diffusion constant. Since the sum over $m$ runs over all adjacent compartments of the $l^{\text{th}}$ compartment,  the fraction in Equation \eqref{diffusion_rdme_pr} is just the discretised version of the Laplace operator $\Delta = \partial_1^2 + \ldots + \partial_M^2$. Introducing a discretised density field $u(x^l)=u^l/{h^M}$, where $x^l$ is the centre of the $l^{\text{th}}$ compartment, and taking the continuum limit of Equation \eqref{diffusion_rdme_pr}, we get the PDE
\begin{align}\label{diffusion_rdme_pr_cont}
  d u(x,t)
  & =
    D \Delta u(x,t) dt,
\end{align}
which is just the diffusion equation for the field $u(x,t)$. 

Consider next the reaction term of the RDME given in Equation \eqref{rdme2}. Since reactions only occur within compartments, we can treat the compartments independently of each other. For a single compartment, Equation \eqref{rdme2} then reduces to the chemical master equation given in Equation \eqref{cme}.
Here, however, we first apply the approximations defined in the Results section to bimolecular reactions and reactions with two non-identical product molecules (see Appendix \ref{app_cox} for more details). These approximations lead to a real-valued PR and only reactions with two identical product molecules lead to stochastic terms. The PR Langevin equation hence simplifies to 
\begin{align}\label{rct_rdme_pr_cont}
  d u_i
  & = 
    \sum_{r=1}^R S_{ir} g_r(\mathbf{u}) dt 
    + \sum_{r'} \sqrt {2 g_{r'}(\mathbf{u})} dW_{r'},
\end{align}
where $\mathbf{u}=(u_1, \ldots, u_N)$ and the sum over $r'$ runs over all reactions with two product particles of species $X_i$. The propensities $g_r(\mathbf{u})$ are obtained by replacing the $n_i$ variables with $u_i$ variables and $\Omega$ with $h^M$ in the expressions for the $f_r$ propensities of the approximated reactions. The factor of two in the square root in Equation \eqref{rct_rdme_pr_cont} comes from the fact that two identical  molecules become produced in these reactions. Reintroducing the label $l$ denoting the compartment number in Equation \eqref{rct_rdme_pr_cont}, and the species label $i$ in Equation \eqref{diffusion_rdme_pr_cont}, we can add the two contributions to obtain
\begin{equation}\label{rdme_pr}
\begin{split}
  d u_i^l 
  & =
    D_i \frac{2 M u_i^l - \sum_{m \in \mathcal{N}(l)} u_i^m}{h^2} dt
    +\sum_{r=1}^R S_{ir} g_r(\mathbf{u}^l) dt \\
  & \quad \quad
    + \sum_{r'} \sqrt {2 g_{r'}(\mathbf{u}^l)} dW_{r'}^l,
\end{split}
\end{equation}
where $\mathbf{u}=(u_1^1, \ldots, u_N^1, \ldots, u_1^L, \ldots, u_N^L)$ and $u_i^l$ is the PR variable of species $X_i$ in the $l^{\text{th}}$ compartment.
If we again define discretised density fields $u_i(x^l)=u_i^l/{h^M}$, where $x^l$ is the centre of the $l^{\text{th}}$ compartment, and $dW_r(x^l) = dW_r^l / \sqrt{h^M}$, we can take the continuum limit of Equation \eqref{rdme_pr} which leads to the real-valued SPDE for the intensity fields given in Equation \eqref{pp_eq1}.
The $g_r(\mathbf{u}(x,t))$ therein are not functions of single PR variables anymore, but rather functionals of the space-dependent intensity field vector $\mathbf{u}(x,t)=(u_1(x,t), \ldots, u_N(x,t))$. They are obtained by taking the corresponding propensity functions $f_r(\mathbf{n})$ of the approximate reactions in real space, replacing $n_i \to u_i(x,t)$ and $\langle n_i \rangle \to \langle u_i(x,t) \rangle$ and omitting $\Omega$ factors. The latter can be identified with $h^M$ here and hence get absorbed in the definition of the intensity fields given below Eq.~\eqref{rdme_pr}.

As an example, consider the reaction $A + B \to \varnothing$. The non-spatial propensity in real space for this reaction is $f(n_A, n_B) = k n_A n_B / \Omega$. However, since this is a bimolecular reaction and hence leads to a complex-valued PR, we replace it by the two reactions $A \to \varnothing$ and $B \to \varnothing$ with propensities $f(n_A, n_B) = k \langle n_B \rangle n_A  / \Omega$ and $f(n_A, n_B) = k \langle n_A \rangle n_B / \Omega$, respectively. By replacing $n_i \to u_i(x,t)$ and $\langle n_i \rangle \to \langle u_i(x,t) \rangle$ and omitting $\Omega$ terms, we thus obtain the corresponding propensity functions in spatial PR space as $g_1(u_A(x,t),u_A(x,t)) = k \langle u_B(x,t) \rangle u_A(x,t)$ and $g_2(u_A(x,t),u_A(x,t)) = k \langle u_A(x,t) \rangle u_B(x,t)$, respectively. See Appendix \ref{app_cox} for more details and examples. \\

\noindent
\textbf{Poisson and Cox processes.}
A (spatial) Poisson process on a spatial region $\mathcal{D}$ of arbitrary dimension defines a measure on countable unions of zero-dimensional subsets (points) of $\mathcal{D}$. A Poisson process is often characterised by an intensity function $u: \mathcal{D} \to \mathbb{R}^+$ giving the probability density of finding a point in an infinitesimal region around $x$. Now let $N(A)$ denote the number of points in a subregion $A \subset \mathcal{D}$. Then $N(A)$ is a Poisson random variable with mean given by the integral of $u(\cdot)$ over $A$: 
\begin{align}
  & p(N(A)=n)
   = 
    \mathcal{P} (n ; u_A), \quad
   u_A
   =
    \int_{A} dx~u(x),
\end{align}
where $\mathcal{P} (n ; u_A)$ is a Poisson distribution in $n$ with mean $u_A$. 

A (spatial) Cox process is a generalisation of a Poisson process and also called ``doubly stochastic process'', in the sense that the intensity function $u$ is itself a random process. Conditioned on the intensity $u$, the Cox process reduces to a Poisson process. The distribution of the number of points in a subregion $A \subset \mathcal{D}$ is hence a mixture of Poisson distributions, 
\begin{align}
  p(N(A)=n)
  & = 
    \int d u_A  \mathcal{P} (n ; u_A) p(u_A).
\end{align}
Since we are interested in dynamical systems, we will assume time-dependent intensities $u: \mathcal{D} \times \mathcal{T} \to \mathbb{R}^+$, where $\mathcal{T}$ is a finite real interval denoting time. We then require that for any fixed time point $t\in \mathcal{T}$ the process is a spatial Poisson (Cox) process with intensity $u(\cdot, t)$. In the case of a Poisson (Cox) process, the intensity $u$ may for example be defined as the solution of a PDE (SPDE).

\section*{Acknowledgments}
This work was supported by the Biotechnology and Biological Sciences Research Council  [BB/F017073/1]; the Leverhulme Trust [RPG-2013-171]; and the European Research Council [MLCS 306999].
The authors thank Peter Swain, Andrew Zammit-Mangion, Manfred Opper and Giulio Caravagna for valuable discussions and comments on the draft.

\begin{appendix}

\section*{\centering \huge Appendix}
\section{Cox process representation and mean-field approximations}\label{app_cox}

\subsection{Classification of reactions w.r.t. to their Poisson representation}\label{app_classification}

As mentioned in the Methods section, the Poisson representation (PR) becomes complex depending on the 
reactions in the system.  Table \ref{tab_pr_tax} shows a classification of different types of elementary reactions in terms of the behaviour of the corresponding diffusion matrix $\mathbf{B}(\mathbf{u})$. We note that this strict classification of course only holds if the considered reaction is the only reaction in the system. If there are several reactions happening, the system typically behaves as the entry in  Table \ref{tab_pr_tax} corresponding to the reaction of highest type.

The behaviours of the PR are quite intuitive: for reactions of Type I, it is well-known that  fluctuations are Poissonian, which manifests itself in a deterministic PR. Note that if the Poisson representation is real-valued, the probability distribution of the molecule numbers in the PR ansatz given in Equation (3) in the main text is a real-valued mixture of Poisson distributions, for which it is well-known that the resulting distributions are super-Poissonian. 
Reactions of Type II, for which fluctuations are super-Poissonian, therefore have real and stochastic PRs. It is easy to see that reactions of Type III and IV, however, cannot be represented in this way: a zeroth or first order reaction with two \emph{non-identical} product molecules, i.e., of Type III, imposes a constraint on the particle numbers. For the reaction $\varnothing \rightarrow A + B$ for example, the particle numbers of species $A$ and $B$ differ by a constant integer number (depending on the initial condition). Conditioned on the molecule number of $A$, $B$ is a delta distribution, which clearly cannot be achieved by a real Poisson mixture, and the PR has to be complex. Bimolecular reactions give rise to similar constraints or may lead to sub-Poissonian fluctuations, and hence their PR has to be complex-valued. We therefore approximate reactions of Type III and Type IV as described in the following.

\subsection{Approximation of Type III and IV reactions}\label{sec_cox_linearisation}

\begin{table*}[t]
\caption{\textbf{Classification of different types of reactions w.r.t. to their Poisson representation.} If different types of reactions are happening, the PR typically behave like the reaction of highest type. \vspace{0.2cm}}
\centering
\begin{tabular}{|c|c|c|c||c|}

\multicolumn4c{reaction types}&
\multicolumn1c{PR}
\cr
\hline
  Type & stoichiometry & description &  examples  &      \cr \hline 
  I    & 
 \pbox{10cm}{$\sum_i s_{ir} \leq 1$ \\$ \sum_i r_{ir} \leq 1$}   
 &  \pbox{10cm}{$ $ \\ zero or one  \\  reactant and  \\ product  molecules \\ $ $}   &   \pbox{10cm}{$\varnothing \rightarrow A $  \\ 
 $A \rightarrow \varnothing$ \\ $A \rightarrow B$}  & real and determ. 
 \cr \hline
  II & \pbox{10cm}{$\sum_i s_{ir} \leq 1$ \\$ r_{ir} = 2$ for one $i$ \\ and  zero otherwise}   &  \pbox{10cm}{$ $ \\ zero or one reactant;  \\ two \emph{identical} \\ product  molecules \\ $ $}   &   \pbox{10cm}{$\varnothing \rightarrow A+A $ \\ 
 $A \rightarrow A+A$ \\ $B \rightarrow A+A$}  & real and stoch. 
 \cr \hline
   III & \pbox{10cm}{$\sum_i s_{ir} \leq 1$ \\$ r_{ir}=r_{jr} = 1$ for two $i \neq j$ \\ and zero otherwise} & \pbox{10cm}{$ $ \\ zero or one reactant; \\two \emph{non-identical}  \\  product molecules \\ $ $}     &   
   \pbox{10cm}{$\varnothing \rightarrow A+B $ \\ $A \rightarrow A+B$ \\ $A \rightarrow B+C$}  & complex and stoch. 
 \cr \hline 
    IV & \pbox{10cm}{$ $ \\ $\sum_i s_{ir} = 2$ \\$ \sum_i r_{ir} \leq 2$  \\ $ $} & two reactant molecules   &   
    \pbox{10cm}{$A+A \rightarrow \ldots $ \\  $A+B \rightarrow \ldots$}  & complex and stoch. 
 \cr \hline 
\end{tabular}\label{tab_pr_tax}
\end{table*}

We would like a real PR for general reaction networks. We therefore have to approximate reactions of Type III and IV. Consider first reactions of Type IV, where two molecules react with each other. We approximate this type of reactions in a \emph{mean-field} type of sense: we replace the interaction of the two molecules with each other by two unimolecular reactions whose propensity functions depend on the particle number of one species and the \emph{mean value} of the respective other species. For instance, the reaction
\begin{align}\label{biom1}
  A + B \xrightarrow{\quad k \quad } \varnothing,   \quad f(\mathbf{n}) = \frac{k}{\Omega} n_B  n_A,
\end{align}
becomes replaced by the two reactions
\begin{align}\label{linear1}
  A \xrightarrow{\quad k \langle n_B \rangle / \Omega \quad } \varnothing,  \quad f(\mathbf{n}) = \frac{k}{\Omega} \langle n_B \rangle n_A, \\
  \label{linear2}
  B \xrightarrow{\quad k \langle n_A \rangle / \Omega \quad } \varnothing,  \quad f(\mathbf{n}) = \frac{k}{\Omega} \langle n_A \rangle n_B,
\end{align}
where $ \langle n_A \rangle$ and $ \langle n_B \rangle$ denote the mean values of the molecule numbers of species $A$ and $B$, respectively, and $\Omega$ is the volume of the system. The reactions \eqref{linear1} and \eqref{linear2} thus correspond to linear reactions with one reactant and zero product molecules. The corresponding PR is therefore real and deterministic. Since the mean values of the corresponding PR variables, say $u_{A}$ and $u_{B}$, are equal to the means of $\langle n_{A} \rangle$ or $\langle n_{B} \rangle$, the rate constants in PR space simply become rescaled by $\langle u_{A} \rangle / \Omega$ and $\langle u_{B} \rangle / \Omega$, respectively. Specifically, if there are no other reactions happening in the system, the PR Langevin equations read
\begin{align}
  du_A 
  & =
    - \frac{k}{\Omega} \langle u_B \rangle u_A dt, \\
  du_B
  & =
    - \frac{k}{\Omega} \langle u_A \rangle u_B dt.
\end{align}
Consider now a bimolecular reaction with two identical reactant molecules,
\begin{align}
  A + A \xrightarrow{\quad k \quad } \varnothing,   \quad f(\mathbf{n}) = \frac{k}{\Omega} n_A (n_A-1). 
\end{align}
For such reactions, we replace the interaction of $A$ with itself by the interaction of $A$ with its mean,
\begin{align}
  A \xrightarrow{\quad k \langle n_A \rangle / \Omega \quad } \varnothing,   \quad f(\mathbf{n}) = \frac{k}{\Omega} \langle n_A \rangle n_A. 
\end{align}
In PR space, this leads to the Langevin equation for $A$,
\begin{align}
  du_A 
  & =
    - \frac{k}{\Omega} \langle u_A \rangle u_A dt.
\end{align}
Consider next the following reaction of Type III (c.f. Table \ref{tab_pr_tax})
\begin{align}
  A \xrightarrow{\quad k  \quad } A + B ,   \quad f(\mathbf{n}) = k  n_A ,
\end{align}
which can be approximated in a similar fashion as the bimolecular reactions before: we replace the dependence of the creation of $B$ molecules on $A$ molecules by a dependence on the mean of the later, i.e.~$\langle n_A \rangle$,
\begin{align}
  \varnothing \xrightarrow{\quad k \langle n_A \rangle  \quad } B ,   \quad f(\mathbf{n}) = k \langle n_A \rangle.
\end{align}
For the other two Type III reactions,
\begin{align}\label{type3_rct2}
  & \varnothing \xrightarrow{\quad k  \quad } B + C ,   \quad f(\mathbf{n}) = k \Omega, \\
  & A \xrightarrow{\quad k  \quad } B + C ,   \quad f(\mathbf{n}) = k n_A,
\end{align}
we have to decouple the productions of $B$ and $C$, which can be achieved by approximating the reactions by  
\begin{align}\label{type3_rct2_approx}
  & \varnothing \xrightarrow{\quad k  \quad} B, \quad \varnothing \xrightarrow{\quad k  \quad } C ,   \quad f(\mathbf{n}) = k \Omega, \\
  & \label{type3_rct2_approx2}
  A \xrightarrow{\quad k  \quad} B, \quad A \xrightarrow{\quad k  \quad} C ,   \quad f(\mathbf{n}) = k n_A.
\end{align}
While \eqref{type3_rct2} correlates the molecule numbers of species $B$ and $C$, we have effectively decorrelated $B$ and $C$ by introducing the reactions \eqref{type3_rct2_approx} and \eqref{type3_rct2_approx}.

Table \ref{tab_approx_rct} summarises the approximations for all reactions of Type III and IV. Note that bimolecular reactions (Type IV) with two identical product molecules under these approximations still lead to stochastic PRs. Note also that depending on the reaction, a combination of the used approximations has to be performed, for example for the reactions $A+B \rightarrow A+C$ or $A+A \rightarrow C+D$.

\subsubsection*{Example}

As an example, consider the following reaction system
\begin{align}\label{pr_space_example_system}
  X \xrightarrow{\quad k_1  \quad} X + X ,  \quad  X + X \xrightarrow{\quad k_2  \quad} \varnothing.
\end{align}
The corresponding stoichiometric matrix reads
\begin{align}
  \mathbf{S} = (1, -2). 
\end{align}
The first reaction in \eqref{pr_space_example_system} is of Type II and thus does not need to be approximated. The corresponding non-spatial propensity function in real space is given by $f_1(n) = k_1 n$, where $n$ is the variable denoting the number of $X$ particles. The second reaction in Eq.~\eqref{pr_space_example_system} is of Type IV and hence needs to be approximated. According to  Table \ref{tab_approx_rct} we approximate it by the reaction $X \xrightarrow{\quad k_2 \langle n \rangle / \Omega  \quad} \varnothing$ with propensity $f_2(n) = k_2 \langle n \rangle n / \Omega$. The corresponding propensity functions in spatial PR space are obtained by replacing $n \to u(x,t)$ and $\langle n \rangle \to \langle u(x,t) \rangle$, where $u(x,t)$ is the PR field of species $X$. We thus have  
\begin{equation}
\begin{split}
  X \xrightarrow{\quad k_1  \quad} X & + X , \quad f_1(n) = k_1 n, \\
  & \downarrow \\
   X \xrightarrow{\quad k_1  \quad} X & + X, \quad g_1(u(x,t)) = k_1 u(x,t), 
\end{split}
\end{equation}
for the first reaction and 
\begin{equation}
\begin{split}
  X + X \xrightarrow{\quad k_2  \quad} \varnothing &  , \quad f_2(n) = \frac{k_2}{\Omega} n(n-1), \\
  & \downarrow \\
   X \xrightarrow{\quad k_2  \quad} \varnothing & , \quad f_2 (n) = \frac{k_2}{\Omega} \langle n \rangle  n, \\
  & \downarrow \\
   X \xrightarrow{\quad k_2  \quad} \varnothing & , \quad g_2(u(x,t)) = k_2 \langle u(x,t) \rangle  u(x,t), 
\end{split}
\end{equation}
for the second reaction. The corresponding stoichiometric matrix becomes 
\begin{align}
  \mathbf{S} = (1, -1). 
\end{align}
Using the general stochastic partial differential equation (SPDE )for intensity fields given in (4) in the main text we hence obtain the for the intensity field $u(x,t)$, 
\begin{equation}
\begin{split}
  d u(x,t) 
  & = 
    [D \Delta u(x,t) + k_1 u(x,t) - k_2 \langle u(x,t) \rangle  u(x,t)] dt \\
 &   \quad
    + \sqrt {2 k_2 \langle u(x,t) \rangle  u(x,t)} dW(x,t).
\end{split}
\end{equation}
We would like to emphasise that $\langle u(x,t) \rangle$ denotes the local expectation of the stochastic intensity field $u(x,t)$ and not a spatial averaging.

\begin{table}[t]
\caption{\textbf{Reactions of Types III and IV and  their approximate reactions.} The corresponding propensities in PR space for the approximate system are obtained by replacing $n_A$ and $n_B$ with $u_A$ and $u_B$, respectively. $n_A$ and $n_B$ denote the particle number variables of species $A$ and $B$, respectively, and $u_A$ and $u_B$ the corresponding PR variables. \vspace{0.2cm}}
\centering
\begin{tabular}{|c|c||c|c|}
\multicolumn2c{reactions}&
\multicolumn2c{approximation}
\cr
\hline
  type & propensity & type &  propensity     
   \cr \hline
 $A+B \rightarrow \ldots$ & $k n_A n_B / \Omega$ &  \pbox{10cm}{$A \rightarrow \ldots$ \\ $ B \rightarrow \ldots$}   &  \pbox{10cm}{$k  \langle n_B \rangle n_A / \Omega $ \\  $ k  \langle n_A \rangle n_B / \Omega $} 
 \cr \hline
  $A+A \rightarrow \ldots$ & $k n_A (n_A-1) / \Omega$ &  $A \rightarrow \ldots$   &$k  \langle n_A \rangle n_A / \Omega $
 \cr \hline
  $A \rightarrow A+B$ & $k n_A $ &  $\varnothing \rightarrow B$   &$k  \langle n_A \rangle $
 \cr \hline
   $\varnothing \rightarrow B+C$ & $\Omega k $ &  \pbox{10cm}{$\varnothing \rightarrow B$ \\ $ \varnothing \rightarrow C$}   & \pbox{10cm}{$\Omega k $ \\  $ \Omega k $}
 \cr \hline
    $A \rightarrow B+C$ & $k n_A $ &  \pbox{10cm}{$A \rightarrow B$ \\ $ A \rightarrow C$}   & \pbox{10cm}{$k n_A $ \\  $ k n_A $}
 \cr \hline
\end{tabular}\label{tab_approx_rct}
\end{table}
%

\subsection{Proof of Remark 1 and Result 1}\label{app_proof}

The proof of Remark 1 and Result 1 relies on the SPDE in Equation (4) in the main text and its derivation given in the Methods section. For simplicity, we consider a one-dimensional system with one species $X$ in the interval $[0,1]$ here. Consider the PR of the RDME for approximated reactions given in Equation (29). 
Consider first a system involving only reactions of Type I. In that case we do not have to perform any approximations to obtain Equation (29) and the second sum including the noise terms vanishes, i.e., Equation (29) reduces to a PDE. 
 For deterministic initial conditions the $u_i$ thus remain deterministic, and the probability distribution of $n^l$ in compartment $l$ at time $t$ is given by a Poisson distribution with mean value $u^l(t)$. The mean number of molecules in an interval $I=[(m_ 1-\frac{1}{2}) h,(m_2+\frac{1}{2})h], m_1<m_2 \in \mathbb{N}$ at time $t$ is thus 
\begin{align}
  \langle N(I, t) \rangle
  & = 
    \sum_{i=m_ 1}^{m_ 2} \langle n^i \rangle  
  = 
    \sum_{i=m_ 1}^{m_ 2} \langle u^i \rangle 
  = 
    \sum_{i=m_ 1}^{m_ 2} u^i,
\end{align}
where $N(I, t) = \sum_{i=m_1}^{m_2} n^i$. Since the $n_i$ are independent Poisson random variables, $N(I,t)$ is also a Poisson random variable with mean $\langle N(I, t) \rangle = \sum_{i=m_ 1}^{m_ 2} u^i(t)$. 

Defining $u^i / h \to u(x^i)$, where $x^l$ is the center of compartment $l$, allows us to take the continuum limit $h \to 0$ of Equation (29)  which gives the (S)PDE in Equation (4).
The mean value of $N(I,t)$ can be written as $\langle N(I, t) \rangle = \sum_{i=m_1}^{m_2} h u(x^i, t)$, which is a Riemann sum. Taking the limit $h \to 0$ for constant $I$ gives
\begin{align}
  \langle N(I, t) \rangle 
  & \to 
    \int_{I} dx u(x,t).
\end{align}
According to the \emph{Countable additivity theorem}, the sum of an infinite number of Poisson distributed independent random variables converges with probability $1$ if the sum of the mean values converges, and the sum has is Poisson distributed with corresponding mean value. We assume that the mean particle density is bound everywhere, which means that the values $u^i /h = u(x^i)$ are bound for all $i$ and all $h$. Let $B$ be such an upper bound. Since 
\begin{align}
  \left| \sum_{i=m_1}^{m_2} h u(x^i) \right|
  & \leq
    h \sum_{i=m_1}^{m_2}  \left| u(x^i) \right|
  \leq
     h \sum_{i=m_1}^{m_2}  B
  = 
    (m_2-m_1)B,
\end{align}
the sum converges in the limit $h \to 0$ for constant $I=[(m_ 1-\frac{1}{2}) h,(m_2+\frac{1}{2})h]$.
We thus find that $N(I,t)$ is Poisson distributed in the continuum limit and we can write 
\begin{align}
  P( N(I, t) = n)
  & \xrightarrow{\quad h \to 0  \quad}    
    \mathcal{P}(n; \int_I dx u(x,t)).
\end{align}
The same can be shown similarly for a countable union of subintervals of $[0,1]$, and $N(U_1,t)$ and $N(U_2,t)$ are obviously independent for disjunct $U_1$ and $U_2$.  The probability distribution for any fixed $t$ is thus \emph{exactly the same as the one of a spatial Poisson process} with intensity $u(x,t)$. 

Suppose now the system also includes reactions of Type II. In this case the PR becomes stochastic, i.e., Equation (29) and its continuum version (4) contain  non-vanishing noise terms.
The field $u(x,t)$ is thus a random process. Given a realisation of $u(x,t)$, the same considerations as for the deterministic case apply and the single-time probability distribution behaves like a spatio-temporal Poisson process. Since $u(x,t)$ is now a random process, the single-time probability distribution of the system \emph{corresponds exactly} to the one of a \emph{spatial Cox process} with intensity $u(x,t)$. The same considerations hold in an approximate sense for Type III and IV reactions.
These findings can easily be generalised to multiple-species systems and general spatial dimensions. This concludes the proof of Remark 1 and Result 1 in the main text.

\section{Inference for Poisson and Cox processes}\label{app_inference}

\subsection{Numerical solution of (S)PDEs via basis projection}\label{sec_basis_proj}

\subsubsection*{General formulation}

To apply the derived Cox process representation we need to solve (S)PDEs. We do this here approximately by  means of a basis function projection leading to a finite set of coupled (stochastic) ordinary differential equations (SDEs/ODEs).
For illustration we confine ourselves here to a one-dimension and one-species system, but the equations can be easily extended to multi-dimensional and multi-species systems. Consider an SPDE of the form
\begin{align}\label{basis_proj_spde}
  d u(x,t) 
  & =
    A(x,t) +\sqrt{C(x,t)} dW(x,t), 
\end{align}
where $A(x,t)$ and $C(x,t)$ are polynomials in $u(x,t)$ with potentially space-dependent coefficients.
We approximate $u(x,t)$ by a linear-combination of a finite set of spatial basis functions $\{\phi_i(x)\}_{i=1}^n$,
\begin{align}\label{basis_proj_ansatz}
  u(x,t)
  & =
    \sum_{i=1}^n c_i(t) \phi_i(x),
\end{align}
where we have introduced the time-dependent coefficients $c_i(t)$.
Inserting this ansatz into \eqref{basis_proj_spde}, multiplying from the left with $\phi_j$ and integrating over $x$, it can be shown that the parameter vector $\mathbf{c}=(c_1, \ldots c_n)$ fulfils
\begin{align}\label{basis_proj_sde}
  d \mathbf{c} (t)
  & =
    \Phi^{-1} \langle \phi | A \rangle dt + \sqrt{\Phi^{-1}  \langle \phi | C | \phi \rangle \Phi^{-1} } d\mathbf{W}, 
\end{align}
where $d\mathbf{W}$ is a $n$-dimensional Wiener process and we have defined
\begin{align}
  \langle \phi_i | f \rangle
  & =
    \int dx \phi_i(x) f(x,t), \\
  \langle \phi_i | f | \phi_j \rangle
  & =
    \int dx \phi_i(x) f(x,t) \phi_j(x), \\
  \langle \phi | f \rangle_i
  & =
     \langle \phi_i | f \rangle, \\
  \langle \phi | f | \phi \rangle_{ij}
  & =
     \langle \phi_i | f | \phi_j \rangle, \\
  \Phi_{ij} 
  & = 
    \langle \phi_i |  \phi_j \rangle, 
\end{align}
for a general function $f(x,t)$.

\subsubsection*{For the real-valued Poisson representation}\label{sec_lin_poiss}

Due to the approximations of certain reaction types introduced in Section \ref{sec_cox_linearisation} the drift and diffusion terms in the SDE in \eqref{basis_proj_sde} are always linear in the coefficient vector $\mathbf{c}$, with coefficients of the drift potentially depending on $\langle \mathbf{c} \rangle$, i.e., the drift may contain terms of the form $c_i \langle c_j \rangle$. Using this it is straightforward to show that the moment equations of $\mathbf{c}$ of different orders are not coupled to each other, i.e., the first-order moment equations depend only on first-order moments, etc. This in turn allows to directly numerically integrate the moment equations. Depending on the reactions involved, the diffusion term may be independent of $\mathbf{c}$ in which case the SDE in \eqref{basis_proj_sde} has a multivariate Gaussian solution. The latter can be obtained by integrating the moment equations of up to order two. If the solution of the SDE is not Gaussian, we simply approximate it by a multivariate Gaussian with mean and variance obtained in the same way. Therefore, with the approximations introduced in Section \ref{sec_cox_linearisation}, the SPDEs of all possible reaction systems can be solved by numerical solution of ODEs without the need for any additional approximations.

\subsubsection*{Locally constant non-overlapping basis functions}

We use locally constant, non-overlapping step functions throughout this work. For a one-dimensional system in the interval $[0,1]$, for example, we define $n$ basis functions as 
\begin{align}
  \psi (x)
  & =  
  \left\{\begin{array}{ll} 
     1  & 0 \leq x \leq \frac{1}{n} ,\\
     0 & \text{otherwise},
  \end{array}
  \right.  \\
  \phi_i(x)
  & =
    \psi (x- (i-1)/n) \quad \text{for} \quad i=1,\ldots n.
\end{align}
The corresponding overlap and diffusion operator matrices read
\begin{align}
  \Phi 
  & =
    \langle \phi  | \phi \rangle
  =
    \frac{1}{n} \mathbb{1}^{n \times n} , \\
  \langle \phi | \Delta | \phi \rangle
  & =
  n
    \begin{pmatrix}
      -1  &  1 &  \\
      1  &  -2 & 1  \\
        &  1  &  -2 &  \ddots \\
        &  &    &  \ddots  &   \\
        &  &  &  \ddots  &  -2  &  1  & \\
        &  &  &  & 1  &  - 2  &  1  & \\
        &  &  &  & &  1  &  -1 \\
    \end{pmatrix},
\end{align}
where $\mathbb{1}^{n \times n}$ is the $n$-dimensional unity matrix and $\Delta$ is the Laplace operator.

\subsection{Filtering}

Here we describe the filtering procedure used to approximate likelihoods.
Consider a Poisson  process with intensity $u(x,t)$ given as the solution of a PDE as in \eqref{basis_proj_spde} (with vanishing noise term),
and spatial measurements $\mathbf{y}=(\mathbf{y}_{t_0}, \ldots, \mathbf{y}_{t_n})$ at discrete times $t_0, \ldots, t_n$. Suppose the intensity is approximated by a linear combination of basis function as in \eqref{basis_proj_ansatz}. Solving the PDE for $u(x,t)$ thus amounts to solving the system of ODEs in \eqref{basis_proj_sde} (with vanishing noise terms) for the coefficient vector $\mathbf{c}$. 

Since the intensity of a Poisson process is deterministic, the likelihood $p(\mathbf{y} | \Theta)$ of the data given the model $\Theta$ is simply computed by solving the ODE in $\mathbf{c}$ forward over the whole time interval and subsequently plugging in the measurements:
\begin{align}\label{poisson_likelihood}
  p(\mathbf{y} | \Theta)
  & =
    \prod_{i=0}^n p(\mathbf{y}_{t_i} | \mathbf{c}(t_i)), \\
\label{poisson_likelihood_single_time}
  p(x_i | \mathbf{c}(t_i))
  & =
     \prod_{s \in x_i} u(s, t_i) e^{- \int dx u(x, t_i)},
\end{align}
where  $u(x, t_i)$ is given in terms of $\mathbf{c}(t_i)$ in \eqref{basis_proj_ansatz}.

In the case of a Cox process, the intensity $u(x,t)$ fulfils an SPDE and thus is a random process. After basis projection as in \eqref{basis_proj_sde} the dynamics can be formulated in terms of the coefficients $c_i(t)$, which fulfil the system of SDEs in \eqref{basis_proj_sde}. As explained in Section \ref{sec_lin_poiss}, the latter is either solved by a Gaussian distribution or we approxiamte it by a Gaussian distribution.
The likelihood has to be computed iteratively by solving the SDEs forward between measurement points and performing measurement updates. Suppose we have the Gaussian posterior $p(\mathbf{c}(t_{i-1}) | \mathbf{y}_{t_{i-1}}, \ldots, \mathbf{y}_{t_0})$ at time $t_{i-1}$. Solving the SDE for $\mathbf{c}$ forward in time we obtain the predictive distribution $p(\mathbf{c}(t_1) | x_{i-1}, \ldots, x_0)$ which is again Gaussian. The posterior at time $t_i$ is then obtained by the Bayesian update as 
\begin{align}\label{bayesian_updat_pp}
  p(\mathbf{c}(t_i) | \mathbf{y}_{t_i}, \ldots, \mathbf{y}_{t_0})
  & =
    \frac{p(\mathbf{y}_{t_i} | \mathbf{c}(t_i)) p(\mathbf{c}( t_i) | \mathbf{y}_{t_{i-1}}, \ldots, \mathbf{y}_{t_0})}{p(\mathbf{y}_{t_i} | \mathbf{y}_{t_{i-1}}, \ldots, \mathbf{y}_{t_0})},
\end{align}
with likelihood contribution
\begin{align}
    p(\mathbf{y}_{t_i} | \mathbf{y}_{t_{i-1}}, \ldots, \mathbf{y}_{t_0})
  & =
     \int d \mathbf{c}( t_i) p(\mathbf{y}_{t_i} | \mathbf{c}( t_i)) p(\mathbf{c}( t_i) | \mathbf{y}_{t_{i-1}}, \ldots, \mathbf{y}_{t_0}),
\end{align}
where $p(\mathbf{y}_{t_i} | \mathbf{c}( t_i))$ is given in \eqref{poisson_likelihood_single_time}. The full likelihood is hence given by
\begin{align}\label{cox_lh}
  p(\mathbf{y} | \Theta)
  & =
    p(\mathbf{y}_{t_0}) \prod_{i=1}^n p(\mathbf{y}_{t_i} | \mathbf{y}_{t_{i-1}},\ldots,\mathbf{y}_{t_0}).
\end{align}
The posterior in \eqref{bayesian_updat_pp} is generally not Gaussian and intractable. We hence approximate it by a Gaussian using the \emph{Laplace approximation}, which approximates the posterior by a Gaussian centred at the posterior's mode and with covariance being the negative Hessian of the posterior in the mode.

\section{Details for studied systems}\label{ch_example_systems}

\subsection{Gene expression}\label{sec_example_gene}

\subsubsection*{Equations}

Consider the gene expression system in Fig.~2 in the main text. For simplicitly, we consider  a one-dimensional version here with the nucleus on one side of the cell. We do not model the gene explicitly, but rather assume a homogeneous production of mRNA in the nucleus. The corresponding reactions are
\begin{align}\label{gene_rct1}
  & \text{nucleus}: \quad \varnothing \xrightarrow{\quad m_1 \quad} M, \\
  \label{gene_rct1b}
  &  \text{whole cell}: \quad M \xrightarrow{\quad m_2 \quad} \varnothing, \\
  \label{gene_rct2}
  & \text{cytosol}: \quad M \xrightarrow{\quad p_1 \quad} M + P, \\
  \label{gene_rct2b}
  & \text{whole cell}: \quad P \xrightarrow{\quad p_2 \quad} \varnothing,
\end{align}
and both the mRNA $M$ and protein $P$ diffuse across the whole cell with diffusion constants $d_M$ and $d_P$, respectively. After approximating the reaction in \eqref{gene_rct2} as explained in Section \ref{sec_cox_linearisation} the PR for this system is real and deterministic, and we obtain using the SPDE in Equation (4)
\begin{align}\label{gene_mrna_pr}
  d u_M(x,t) 
  & = 
    [d_M \Delta u_M(x,t) + m_1 h_n(x) - m_2 u_M(x,t) ] dt, \\
\label{gene_protein_pr}
  d u_P(x,t) 
  & = 
    [d_P \Delta u_P(x,t) + p_1 h_c(x) u_M(x,t) - p_2 u_P(x,t) ] dt, \\
  h_n(x) 
  & =
    \frac{1}{r} \Theta(r-x), \\
  h_c(x)
  & =
    \frac{1}{1-r} \Theta(x-r), 
\end{align}
where $r$ is the size of the nucleus and $\Theta$ the Heaviside step function. The functions $h_n(x)$ and $h_c(x)$ arise because $M$ and $P$ only become created in the nucleus and cytosol, respectively. If we additionally include the autocatalytic reaction 
\begin{align}\label{autocatalytic}
  P \xrightarrow{\quad p_3 \quad} P + P,
\end{align}
the equation for $u_P(x,t)$ becomes an SPDE and reads
\begin{align}\label{gene_auto_protein_pr}
  d u_P(x,t) 
  & = 
    \Big[ d_P \Delta u_P(x,t) + p_1 h_c(x) u_M(x,t) + p_3 u_P(x,t) \\
  & \quad  - p_2 u_P(x,t) \Big] dt + \sqrt{2 p_3 u_P(x,t)} dW(x,t).
\end{align}
%

\subsubsection*{Inference}

Consider first the system without the reaction in \eqref{autocatalytic}. In this case the system corresponds to a Poisson process. After basis function projection of the PDEs in \eqref{gene_mrna_pr} and \eqref{gene_protein_pr} as explained in Section \ref{sec_basis_proj}, we are left with solving a coupled system of ODEs and can compute data likelihoods as in \eqref{poisson_likelihood}.
We fix the parameters to
\begin{equation}\label{gene_parameters}
\begin{split}
  r
   & =
    0.3, \quad
  d_M 
   =
    0.1, \quad
  m_1
   =
    20, \quad
  m_2 
   =
    0.5, \\
  & \quad \quad
  d_P
    =
    0.1, \quad
  p_1
   =
    20, \quad
  p_2
  =
    0.2.
    \end{split}
\end{equation}
We assume that initially there are  zero mRNA molecules and zero protein molecules in the cell. We further assume that the mRNA is unobserved and consider measurements of the protein at thirty equally separated time points separated by $\Delta t = 0.5$. We project the PDEs in \eqref{gene_mrna_pr} and \eqref{gene_protein_pr} onto twenty basis functions as explained in Section \ref{sec_basis_proj}. We then optimise the likelihood of the data with respect to the parameters to obtain parameter estimates. We vary the initial values for the parameters in the likelihood optimiser randomly between $0.5$ times and $2$ times the exact value. The inference results are shown in Table 1 in the main text.

Next, we consider the same system but with the additional reaction in \eqref{autocatalytic}, for which the PDE in \eqref{gene_protein_pr} gets replaced by the SPDE in \eqref{gene_auto_protein_pr}. Now  the system corresponds to a Cox process and we are left with solving a coupled system of SDEs after basis function projection. We approximate the solution of the SDEs by a multivariate Gaussian as described in Section \ref{sec_basis_proj}. The corresponding likelihoods can then be computed as in \eqref{cox_lh}.
We again consider measurements of the protein at equally separated time points separated by $\Delta t = 0.5$ and optimise the corresponding likelihood. The results are shown in Table 2 in the main text.

\subsection{SIRS}\label{sec_example_sirs}

\subsubsection*{Equations}

The reactions of the SIRS system are 
\begin{align}
  S+I \xrightarrow{\quad k, w \quad} 2I, \quad
  I \xrightarrow{\quad r \quad} R,  \quad
  R \xrightarrow{\quad s \quad} S.
\end{align}
We consider a system in the two-dimensional square $[0,1] \times [0,1]$. After approximating the reaction in \eqref{gene_rct2} as explained in Section \ref{sec_cox_linearisation} the PR for this system is real, and we obtain using Equation (4) for the intensity fields of $S$, $I$ and $R$,
\begin{align}
  d u_S(x,t)
  & =
    d \Delta u (x,t) - k^{\text{PR}} u_S(x,t) u_I(x,t)  + s u_R(x,t), \\
  d u_I(x,t)
  & =
    d \Delta u_I (x,t) + k^{\text{PR}} u_S(x,t) u_I(x,t)  - r u_I(x,t), \\
  d u_R(x,t)
  & =
    d \Delta u_R (x,t) + r u_I(x,t)  - s u_R(x,t),
\end{align}
where we omitted noise terms in the equation for $u_I(x,t)$ for simplicitly and hence treat the system deterministically. We introduced the reaction rate $k^{\text{PR}}$ in the term corresponding to the bimolecular infection reaction.
If we include the additional spontaneous infection reaction
\begin{align}
  S \xrightarrow{\quad v \quad} I,
\end{align}
the equations for $u_S(x,t)$ and $u_I(x,t)$ obtain an additional term and read
\begin{align}
  d u_S(x,t)
  & =
    d \Delta u (x,t) - k^{\text{PR}} u_S(x,t) u_I(x,t)  + s u_R(x,t) - v u_S(x,t), \\
  d u_I(x,t)
  & =
    d \Delta u_I (x,t) + k^{\text{PR}} u_S(x,t) u_I(x,t)  - r u_I(x,t) + v u_S(x,t).
\end{align}
%

\subsubsection*{Inference}

As an initial condition we distribute $S_{ini}$ particles of species $S$ randomly across the whole area, one $I$ particle at $[0.05,0.05]$ and assume zero $R$ particles. We simulate data for forty time points equally spaced by $\Delta t = 1$. As a basis we take $100$ basis functions equally distributed in both dimensions. The inference results are shown in Table 3 in the main text.

\subsection{Drosophila embryo}\label{sec_example_drosophila}

\subsubsection*{Data and equations}

The data of the Bicoid protein in Drosophila embryos used here consists of two-dimensional fluorescence data as depicted in Fig.~4a in the main text. Since the relation of measured fluorescence intensity to 
actual protein numbers is unknown we simply translate them one to one here. 
The Bicoid is typically modelled by a simple birth-death process with the reactions
\begin{align}
  \varnothing \xrightarrow{\quad k_1 \quad} P, \quad P \xrightarrow{\quad k_2 \quad} \varnothing.
\end{align}
For simplicity, since diffusion is radially symmetric, we only consider the data within a certain distance from the major axis of the embryos, thus effectively obtaining one-dimensional data. We assume further that the protein is produced within a certain range around the left tip of the embryos. Mathematically the system is thus equivalent to the mRNA system in Section \ref{sec_example_gene}.
The intensity of the protein hence fulfils the PDE 
\begin{align}\label{drosophila_pde}
  d u(x,t)
  & =
    (d \Delta u(x,t) + k_1 f(x) - k_2 u(x,t)) dt,
\end{align}
where $x$ is the distance from the left end of the embryo, $d$ is the diffusion constant, $k_1$ the production rate, $f(x)=1, x \in [0,r], f(x)=0, x \notin [0,1]$, $r$ is the production radius around the origin and $k_2$ is the decay rate.

\subsubsection*{Inference}

Since we have steady-state data, not all parameters are identifiable. 
One can easily see that multiplication of $k_1$, and $k_2$ with the same factor leads to the same steady state. We thus infer the creation range $r$, the diffusion rate $d$, and the ratio $c=k_2/k_1$. For the inference we project the PDE in \eqref{drosophila_pde} on twenty basis functions and solve the resulting ODEs for large times to ensure the solution to be in steady state. We optimise the likelihood for each of the embryos independently to obtain the inferred parameter values. The results are visualised in Fig.~4 in the main text.

\end{appendix}
\end{document}